\documentclass{emulateapj}
\usepackage{url}
\usepackage{gensymb}
\usepackage{textcomp}
\usepackage{xspace}
\usepackage{natbib,amsmath} \usepackage{graphicx}
\usepackage{hyperref} \usepackage{float} \usepackage{subfigure}
\usepackage{tabularx}
\providecommand{\e}[1]{\ensuremath{\times 10^{#1}}}
\newcommand{\um}{\textmu m\xspace}
\newcommand{\emcee}{\texttt{emcee}\xspace}
\newcommand{\spitzer}{\emph{Spitzer}\xspace}
\newcommand{\batman}{\texttt{batman}\xspace}
\newcommand{\methane}{CH\textsubscript{4}\xspace}
\usepackage{array}
\newcolumntype{C}{>{$}c<{$}} % math-mode version of "c" column type
\newcolumntype{R}{>{$}r<{$}} % math-mode version of "r" column type

\begin{document}
\title{Phase curves of WASP-33b and HD 149026b and a New Correlation Between Phase Curve Offset and Irradiation Temperature}

\author{ Michael Zhang\altaffilmark{1}, Heather
  A. Knutson\altaffilmark{2}, Tiffany Kataria\altaffilmark{3}, Joel C.
  Schwartz\altaffilmark{4}, Nicolas B. Cowan\altaffilmark{4}, Adam
  P. Showman\altaffilmark{5}, Adam Burrows\altaffilmark{6}, Jonathan J. Fortney\altaffilmark{7}, Kamen
  Todorov\altaffilmark{9}, Jean-Michel Desert\altaffilmark{9}, Eric Agol\altaffilmark{8}, Drake Deming\altaffilmark{10}}

\altaffiltext{1}{Department of Astronomy, 
	California Institute of Technology, Pasadena, CA 91125, USA; 
	mzzhang2014@gmail.com}
\altaffiltext{2}{Division of Geological and Planetary Sciences,
California Institute of Technology, Pasadena, CA 91125, USA}
\altaffiltext{3}{Jet Propulsion Laboratory, Pasadena, CA 91109, USA}
\altaffiltext{4}{Department of Physics, McGill University, Montreal, Quebec H3A 2T8, Canada}
\altaffiltext{5}{Department of Planetary Sciences and Lunar and Planetary Laboratory, University of Arizona, Tucson, Arizona 85721, USA}
\altaffiltext{6}{Department of Astrophysical Sciences, Princeton University, Princeton, NJ 08544, USA}
\altaffiltext{7}{Department of Astronomy and Astrophysics, University of California at Santa Cruz, Santa Cruz, CA 95604, USA}
\altaffiltext{8}{Department of Astronomy, University of Washington, Seattle, WA 98195, USA}
\altaffiltext{9}{API, University of Amsterdam, P.O. Box 94249, 1090 GE Amsterdam, The Netherlands}
\altaffiltext{10}{Department of Astronomy, University of Maryland, College Park, MD 20742, USA}

\begin{abstract}
  We present new 3.6 and 4.5 \um \spitzer phase curves for the highly irradiated hot Jupiter WASP-33b and the unusually dense
  Saturn-mass planet HD 149026b.  As part of this analysis, we develop a new variant of 
  pixel level decorrelation that is effective at removing intrapixel
  sensitivity variations for long observations ($>$10 hours) where the position of the star can vary by a significant fraction of a pixel.  Using this
  algorithm, we measure eclipse depths, phase amplitudes, and phase
  offsets for both planets at 3.6 \um and 4.5 \um. 
  We use a simple toy model to show that WASP-33b's phase offset, albedo, and heat recirculation efficiency are largely similar to those of other hot Jupiters despite its very high irradiation.  On the other hand, our fits for HD 149026b prefer a very high albedo and an unusually high
  recirculation efficiency.  We also compare our results to
  predictions from general circulation models, and find that while
  neither planet matches the models well, the discrepancies for HD 149026b are especially large.  We speculate that this may be related to its high bulk metallicity, which could lead to enhanced atmospheric opacities and the formation of reflective cloud layers in localized regions of the atmosphere. 
  We then place these two planets in a broader context
  by exploring relationships between the temperatures, albedos,
  heat transport efficiencies, and phase offsets of all planets with published thermal phase
  curves.  We find a striking relationship between
  phase offset and irradiation temperature--the former drops with increasing
  temperature until around 3400 K, and rises thereafter.  Although
  some aspects of this trend are mirrored in the circulation models, there are notable differences that provide important clues for future modeling efforts.
  
\end{abstract}

% #####################################################################

\section{Introduction}
\label{sec:introduction}
The \emph{Spitzer Space Telescope} was designed and constructed prior to the discovery of the first transiting exoplanet, but it
has nevertheless become an important tool in the study of exoplanet
atmospheres.  In particular, the development of techniques to correct
instrumental systematics and derive precise timeseries photometry from \spitzer
data has enabled the first measurements of thermal emission from
a diverse array of exoplanets \citep{seager_deming_review}.  These measurements, in the form of secondary
eclipses and phase curves, allow us to characterize the
temperatures, albedos, heat transport efficiencies, and phase offsets of these planets \citep[e.g.,][]{cowan_agol_2011}.  For planets with observations at
multiple wavelengths we can also constrain their atmospheric
compositions, investigate their vertical pressure-temperature
profiles, and probe the presence of clouds \citep{burrows_2010}.
These \spitzer phase curves provide invaluable information about the
fundamental physical processes that drive the atmospheric circulation
patterns of these tidally-locked planets, and can be compared to
predictions from general circulation models \citep[GCMs; e.g.,][]{heng_showman}.  Although
both models and observations are generally in good agreement on the
dayside emission spectra of hot Jupiters, there are significant
discrepancies in the measured night-side spectra
(e.g. \citealp{hd209458b}) and models that provide a good match to the
measured phase curve amplitude and phase offset in a single bandpass
often have difficulties matching phase curve data for the same planet
at additional wavelengths (i.e. \citealp{hd189733b}).  This suggests
that there are aspects of the atmospheric circulation, cloud
properties, magnetic fields, and chemistry of these planets that are
not adequately captured in current GCMs.  The GCMs we use
in this paper, for example, neglect clouds, magnetohydrodynamics, and
disequilibrium chemistry, although some of these topics have been
investigated in more focused modeling studies (e.g.,
\citealt{cooper_showman_2005, parmentier_2016, rogers_2017}).

In this paper, we examine multi-wavelength phase curve observations for two planets with unusual
characteristics as compared to the broader sample of transiting hot Jupiters.  WASP-33b is a 2.2 M\textsubscript{J} planet with a radius of 1.5 R\textsubscript{J} orbiting a 1.5 M\textsubscript{sun} $\delta$ Scuti star
with a period of 1.22 days \citep{wasp33b_mass}. With an irradiation
temperature ($T_0 = T_{\rm eff}/\sqrt{a_*}$) of 3890 K, this planet is one of the most highly irradiated hot Jupiters currently known.  The star itself has pulsations
at a variety of frequencies, with the dominant mode at 21 d\textsuperscript{-1}.  These
pulsations have an amplitude of roughly 1 mmag, about 1/4 that of the
secondary eclipse depth \citep{von_Essen_2014}. Previous authors have measured broadband thermal emission from WASP-33b's dayside at a variety
of wavelengths, including 0.91 \um \citep{smith_2011}, 1.05 \um
\citep{von_Essen_2015}, 2.14 \um
\citep{deming_2012,de_Mooij_2013}, and \spitzer's 3.6 and
4.5 \um bands
\citep{deming_2012}.  \cite{von_Essen_2015}
summarize these results and combine them to obtain an average dayside
brightness temperature of $3358 \pm 165$ K.  More recently,
\cite{wasp33b_inversion} reported evidence for a temperature inversion in the 1.1--1.6 \um dayside spectrum of this planet, and \citet{nugroho_2017} used a cross-correlation technique to detect TiO in the
  0.62--0.88 \um dayside spectrum with the High-Dispersion Spectrograph on Subaru.  It has long been suggested \citep{hubeny_2003, fortney_2008,
  burrows_2008} that additional opacity from molecules such as
gas-phase TiO and VO could lead to the formation of temperature
inversions in the most highly irradiated atmospheres, and this indeed appears
to be the case for WASP-33b.

Atmospheric circulation models generally predict that more highly
irradiated planets should have larger day-night temperature
contrasts.  According to \cite{showman_guillot}, the day-night
temperature difference can be thought of as
resulting from a competition between the radiative cooling timescale
$\tau_{rad}$ and the timescale of advection by wind, $\tau_{adv}$. Because the radiative timescale decreases much faster with increasing temperature
than the advective timescale, more highly irradiated planets should
have steeper day-night temperature gradients.  \cite{perez_becker_showman} and \cite{komacek_showman}
show that the full picture is more complicated, but the general idea is
still that radiation outcompetes other heat transport mechanisms for
the most highly irradiated planets, causing a larger day-night temperature difference.  The typical pressure at whic incident starlight is absorbed is also important for atmospheric circulation, and the presence of a dayside temperature inversion will therefore also affect the redistribution of energy to the planet's night side \citep{sparc_model,lewis_2014}.

HD 149026b is a 0.36 M\textsubscript{J} planet with a radius of 0.65 R\textsubscript{J}, and
orbits a subgiant G0 IV star of metallicity [Fe/H] = 0.36 with a period of 2.9 days \citep{sato_2005}. Its small radius and
correspondingly large density suggest the presence of a large
heavy-element core.  \cite{hd149_core_mass_estimates} summarize the
many attempts to estimate the mass of this core, concluding that
plausible estimates range from 45-110 $M_\earth$, corresponding to
39-96\% of total mass.  Given the high metallicity of the star and the
planet's large core-mass fraction, this planet seems likely to have
a high atmopsheric metallicity.  Although \cite{hd149026b_transit_spectrum}
analyzed four spectroscopic transit observations with \emph{Hubble}'s NICMOS instrument (1.1--2.0 \um),
the uncertainties from these data were too high to provide useful
constraints on the planet's transmission spectrum. 
\cite{stevenson_2012} subsequently obtained \spitzer secondary eclipse observations
at 3.6, 4.5, 5.8, 8.0, and 16 \um and found a brightness temperature of
$2000 \pm 60$ K at 3.6 \um and $1600-1800$ K at longer wavelengths.
When they fit these data with chemical equilibrium models, they found
that they preferred models with large amounts of CO and
CO\textsubscript{2}, 30$\times$ solar metallicity, no temperature inversion,
and moderate heat redistribution.

\cite{lewis_2010} studied the effect
of metallicity on the warm Neptune GJ436b and found that high
metallicity models had equatorial jets and strong day-night
temperature variations, while lower metallicity models had weak
temperature variations and high latitude jets. By contributing opacity, metals raise the
photosphere to a higher altitude, where atmospheric dynamics are less
important and radiative cooling is more efficient.  For planets with
condensate cloud layers, increasing the atmospheric metallicity also
increases the amount of cloud-forming material and the corresponding
cloud opacity.  Although most circulation models do not currently
include clouds, the presence of spatially inhomogeneous cloud layers
can significantly alter the shape of both optical and infrared phase
curves \citep{parmentier_2016,heng_2013,shporer_2015}.

We describe our new 3.6 and 4.5 \um phase curve observations for
WASP-33b and HD 149026b in Section
\ref{sec:observations} and our analysis of this data in Section
\ref{sec:analysis}.  In Section \ref{sec:discussion} we combine a
simple toy model and more sophisticated GCM simulations for each planet to
interpret these observations and search for patterns in the full
sample of published thermal phase curve observations.  Finally we make
concluding remarks in Section \ref{sec:conclusions}.

\section{Observations}
\label{sec:observations}

\begin{table*}[t]
  \centering
  \caption{\emph{Spitzer} Observation Details}
  \begin{tabular}{c C c c c c c c}
      \hline
      Planet & \lambda (\mu m) & Date (UTC) &
      Duration (h) & Frames & Exposure time (s) &
      PLD Order & Aperture radius (pix) \\
      \hline
      WASP-33b & 3.6 & June 4, 2012 & 37.2 & 311,552 &
      0.36 & 2 & 2.5 \\
      WASP-33b & 4.5 & April 11, 2012 & 37.2 & 311,680 &
      0.36 & 1 & 2.8 \\
      HD 149026b & 3.6 & April 8, 2011 & 81.2\footnote{Including 2.3 h of downlink time}  & 663,104 &
      0.36 & 2 & 2.8 \\
      HD 149026b & 4.5 & April 8, 2011 & 81.6\footnote{Including 2.7 h of downlink time} & 663,104 &
      0.36 & 3 & 2.6 \\
      \hline
  \end{tabular}
  \label{table:observations}
\end{table*}

All observations were taken with the 3.6 and 4.5
\um arrays of the IRAC instrument on \spitzer \citep{irac} during the post-cryogenic (warm) mission.  Start dates, total durations (including downlink time), AORs, and other information about the observations used in this paper are presented in Table
\ref{table:observations}.  Observations were timed to begin before a secondary eclipse and end
after the following secondary eclipse, and all frames were taken
in subarray mode without the now-standard peak-up pointing
optimization \citep{peakup}, which was implemented after these
observations were executed.  Due to data volume constraints the HD 149026b observations required a single downlink break near the middle of each phase curve, resulting in a 2-3 hour gap in coverage.  The WASP-33b observations were executed without any breaks for downlinks.

\section{Analysis}
\label{sec:analysis}
\subsection{Overview}
We extract a photometric timeseries for each phase curve observation
using aperture photometry, then fit the data with a combined
astrophysical and noise model as described in the sections below.

\subsection{Photometry}
Subarray images are 32x32 pixels. We 
estimate and subtract the sky background from each image by excluding
all pixels within a radius of 12 pixels from the star, rejecting
outliers using sigma clipping, and then calculating the 
biweight location of the remaining pixels.  The biweight location is a    
robust and efficient statistic implemented in \texttt{astropy}
\citep{astropy}, and we find that it gives results comparable to
methods used in previous studies \citep{intrapixel_methods}.  For both
planets, the
sky background contributes less than 1\% of the total flux at 3.6 \um
and less than 0.5\% of the total flux at 4.5 \um for our preferred apertures.

We estimate the position of the star in each image 
using an iterative flux-weighted centroiding method with a circular
aperture of radius 3 pixels, and perform
aperture photometry using the \texttt{photutils} module \citep{photutils} from \texttt{astropy}.
We consider apertures with fixed radii ranging from 1.5 to 5.0 pixels
in steps of either 0.1 pixels ($1.5-3.0$ pixels) or 0.5 pixels
($3.0-5.0$ pixels).  

We omit the first 0.1 days of data for each data set, which is normal
procedure for \spitzer analyses \citep{pld_paper} and removes an
obvious ramp at the beginning of the observations.  The HD 149026b
3.6 observations have a downlink gap in the middle, so in addition to
removing 0.1 days of data from the very beginning, we also remove 0.1
days from the post-downlink segment.

\subsection{Instrumental noise model}
The largest flux variations in our raw \spitzer light curves are not
astrophysical, but instead result from well-known intrapixel
sensitivity variations combined with telscope pointing jitter (e.g.,
\citealt{charbonneau_2005,pointing_effects}). 
Although there are several different approaches to correcting for
these effects, pixel level decorrelation (PLD; \citealt{pld_paper}) has
been among the most successful to date in fits to shorter ($<$ 10
h) observations \citep{intrapixel_methods}.  Following the updated definition of PLD in \cite{benneke_2017}, we model the light curve as:
\begin{align}
  L(t) = f(t)(1+m(t-t_0))\left(\sum_{i=1}^9 c_iP_i(t)\right),
\end{align}
where m is the slope, $f(t)$ is the true brightness, $P_i(t)$ are the normalized fluxes in a
$3\times3$ pixel box centered on the position of the star, $c_i$
are nine coefficients giving the relative weight of each pixel, and the final two terms are meant to model temporal variations in sensitivity. In each image, we remove astrophysical flux variations by dividing the individual pixel values by the sum of the flux across all nine pixels. 

\begin{figure}[h]
  \centering \subfigure[3.6 \um] {\includegraphics
    [width=0.5\textwidth]{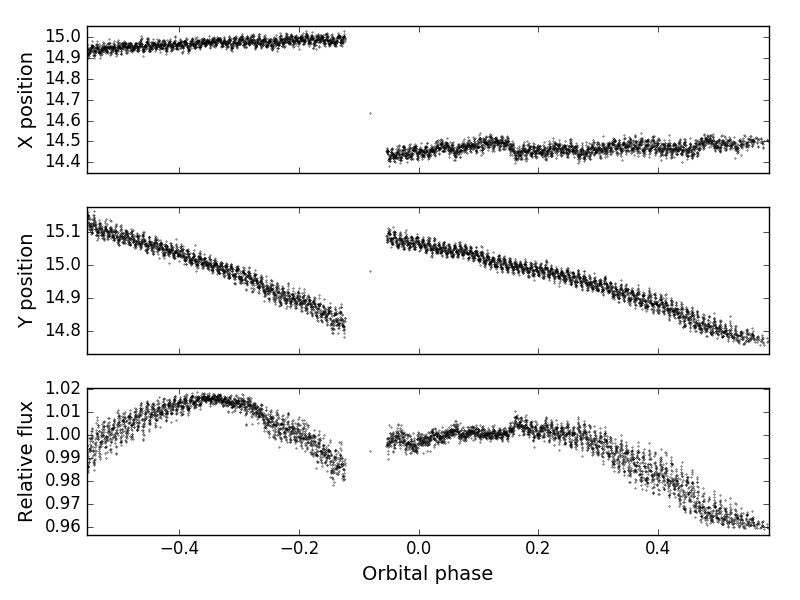}}\qquad
  \subfigure[4.5 \um] {\includegraphics
    [width=0.5\textwidth]{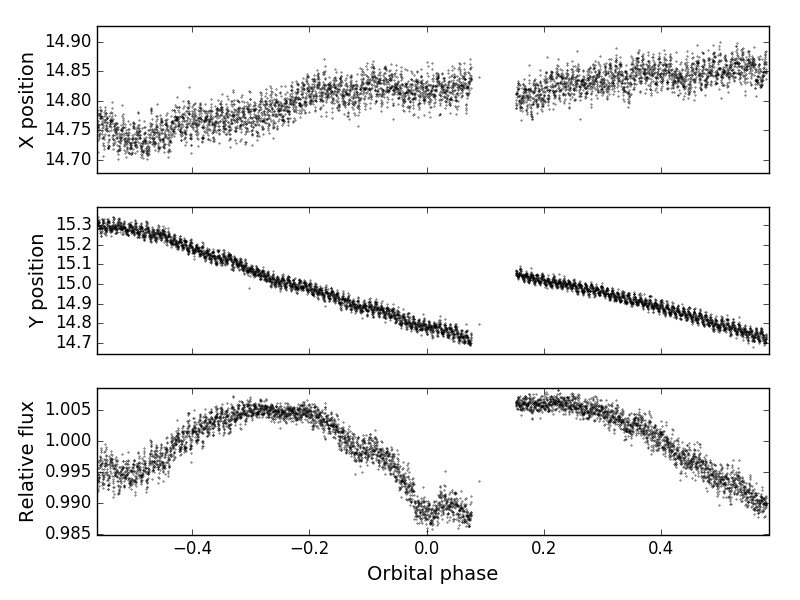}}
    \caption{Raw photometry and $x$ and $y$ position as a function of
      orbital phase for HD 149026b.  The top panel shows the 3.6 \um
      data, and the bottom panel shows the 4.5 \um data.  Fluxes have
      been divided by the median value, and all measurements are shown
      binned into sets of 128 points, corresponding to a time step of
      51 seconds.}
\label{fig:hd149026_raw}
\end{figure}

\begin{figure}[h]
  \centering \subfigure[3.6 \um] {\includegraphics
    [width=0.5\textwidth]{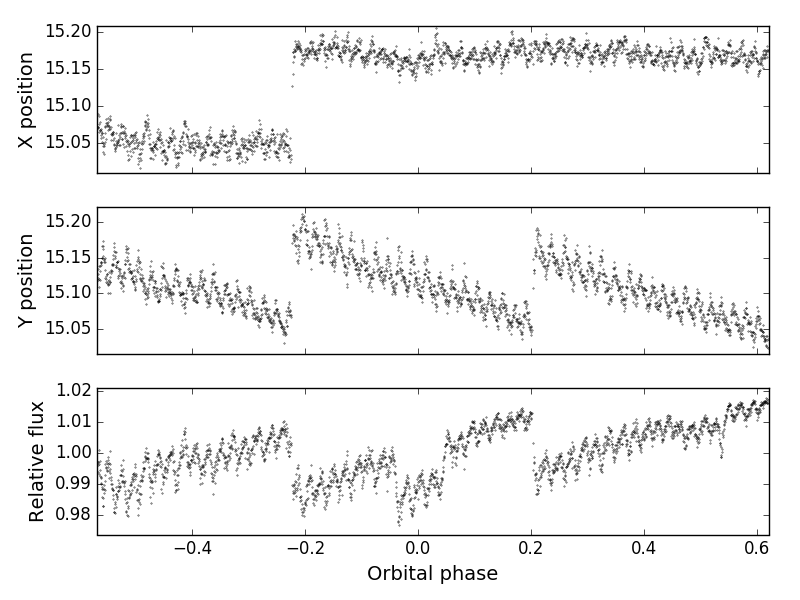}}\qquad
  \subfigure[4.5 \um] {\includegraphics
    [width=0.5\textwidth]{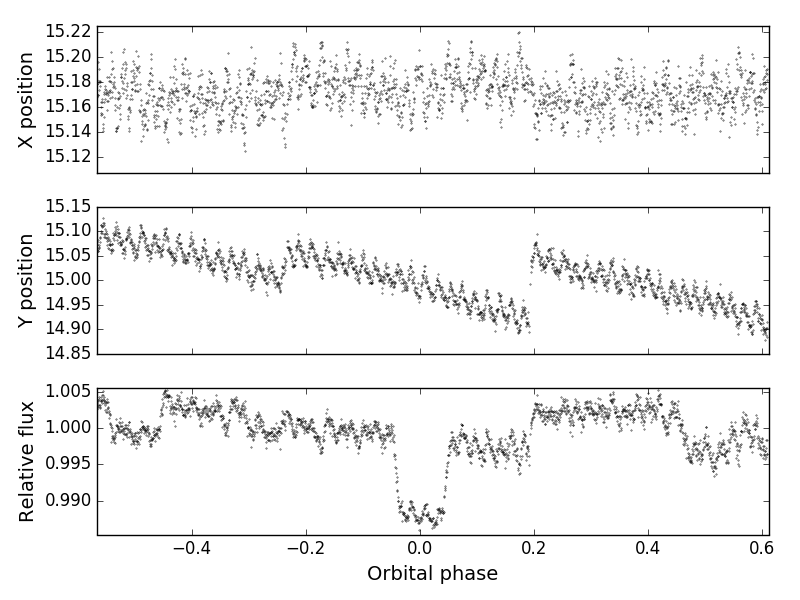}}
    \caption{Raw photometry and $x$ and $y$ position as a function of orbital phase for WASP-33b.  See Fig. \ref{fig:hd149026_raw} caption for additional information.}
\label{fig:wasp33_raw}
\end{figure}

We do not necessarily expect a linear
relationship between individual pixel values and the total flux across
the aperture. PLD was originally formulated as the first term of a
Taylor series expansion and therefore works best when applied to data
where the star moves over a relatively small range (typically on the
order of $1/10^{th}$ of a pixel) of pixel positions \citep{wong_2015}.  In our
observations the star drifted by as much as half a pixel (see Figure
\ref{fig:hd149026_raw}), and we found that the standard linear PLD produced correspondingly poor fits.  We account for this increased drift by developing a new variant of PLD:

\begin{align}
  L(t) = f(t)(1+m(t-t_0))\left(\sum_{i=1}^n \sum_{j=1}^9 c_{ij}P_j(t)^i\right),
  \label{eq:pld}
\end{align}
where the linear slope $m$ is a free parameter and $n$ is the highest
order used in the model.  This is similar to \cite{luger_2016}, except that we
neglect cross terms.  We experimented with cross terms but found that
they did not improve the quality of the fits.  This, combined with the
combinatorial explosion in the number of cross terms as the order is
increased, convinced us to drop the cross terms.

Following \cite{benneke_2017}, we have opted to include the linear term in all our fits.  The inclusion of a linear term is standard in many analyses (e.g., \citealt{stevenson_2012, deming_2015}) and can account for a variety of instrumental and astrophysical noise sources that are not adequately corrected by the basic instrumental noise model.  We find that adding the slope decreases the value of the BIC substantially for WASP-33b's 3.6 \um phase curve ($\Delta BIC=-34.5$) and HD-149026b's 3.6 \um phase curve ($\Delta BIC=-15.3$), while having little effect on the value for HD 149026b 4.5 \um $(\Delta BIC=0.46)$ and increasing it for WASP-33b 4.5 \um ($\Delta BIC=7.7$).  Nevertheless, we include the linear term for all light curves for uniformity.  We also considered a quadratic term but found that it resulted in an increased BIC for all four visits.

We tried fits in which $n$ ranged as high as sixth order, but found that
going beyond third order terms never led to lower BIC. Although we
include the $m(t-t_0)$ term in our instrumental noise model,
following \cite{pld_paper}, it could also represent an
astrophysical drift in the stellar brightness.

Since we use Monte Carlo Markov Chain (MCMC) to fit all parameters,
one challenge is the signficant degree
of degeneracy between individual pixel light curves, which can result
in long convergence times for MCMC fits.  We reduce these degeneracies
and improve convergence times by carrying out a principle component
analysis (PCA) on the $N\times9n$ matrix of central pixel data where N is
the total number of images in each phase curve observation, resulting
in a $N\times9n$ matrix of reprojected central pixel data.

\subsection{Astrophysical model}
Our astrophysical model consists of a transit, a secondary eclipse, and a phase
curve.  To model the transit and eclipse, we developed a
GPU-accelerated version of \batman \citep{batman_package}, which is
roughly 10
times faster than the CPU version.  This code has since been merged with
the main repository\footnote{\url{https://github.com/lkreidberg/batman}}.   When calculating the transit and eclipse shapes
we take the period from \cite{smith_2011} (WASP-33b) and
\cite{hd149026b_transit_spectrum} (HD 149026b) and allow the transit
timing, inclination, transit depth, eclipse depth, eclipse phase
(common to both eclipses), and $a/R_*$ to vary as free
parameters in our fits.
We model the transit using a four-parameter nonlinear limb-darkening
law, with coefficients derived via linear interpolation from
\cite{limb_darkening}.  For WASP-33b, we assumed $T_{\rm eff} = 7400 K$,
log g = 4.3, and [M/H] = 0.1 \citep{wasp33b_discovery}.  For HD
149026b, we assumed $T_{\rm eff} = 6160 K$, log(g)=4.278, and [M/H]=0.36
\citep{torres_2008}.  \batman calculates the eclipse shape from
geometry alone, thus neglecting limb darkening and all other sources of
planetary brightness variation.  Published radial velocity measurements and secondary eclipse times for
both WASP-33b and HD 149026b indicate that the orbital eccentricities
for both of these planets are consistent with zero
\citep{kovacs_2013,von_Essen_2015} and we therefore fix the eccentricities of both planets to zero in our fits.

Following \cite{phase_curve_math}, we model the planet's phase
variation as a series expansion in sine and cosine, where we only
consider first-order sinusoidal terms:

\begin{align}
  L_p = C + c_1\cos(2\pi t/P) + c_2\sin(2\pi t/P),
\end{align}
where P is the orbital period.  Although we also explored fits with second order harmonic terms, we found that these did not improve the quality of the fit for either planet.

\subsection{Noise model}
For HD 149206, which has a relatively quiet host star, we assume the
noise is Gaussian and uncorrelated (i.e., white) and allow the value
of the per-point uncertainty in each bandpass to vary as a free
parameter in our fits.  However, as discussed in \S\ref{sec:introduction},
WASP-33 has quasi-periodic stellar oscillations on the order of
0.1\% that need to be accounted for in order to achieve a good fit.

It is possible to model these oscillations using sinusoidal functions or wavelets, as done by other authors (i.e., \citealt{kovacs_2013,deming_2012,von_Essen_2014}).  However, we decided to use the Gaussian
process code \texttt{celerite} \citep{celerite} 
to fit these pulsations non-parametrically.  Gaussian Processes treats the
pulsations as a form of correlated noise whose properties are
described by a parameterized covariance matrix fitted to our data.  This avoids the need to impose a functional form on the oscillations, and allows the
oscillation modes to depart from perfect periodicity over the course of the
observation.  We also tried using a combination of three sinusoids to fit the stellar pulsations, but found that for the WASP-33b 3.6 \um light curve, this decreased the standard deviation of the residuals by only 10\%, compared to 54\% for the GP code.  Three sinusoids require six free parameters, as compared to the five parameters of our GP model.

\texttt{celerite} models the covariance matrix with a function that depends
only on $\tau = |t_i-t_j|$, the time difference between two observations.
We define the
covariance function as the sum of two radial kernels patterned after a
simple harmonic oscillator along with a diagonal kernel to represent
the white noise (the latter term is functionally the same as the white
noise parameter in our HD 149026 fits).  The kernel $k(\tau)$
representing the correlated noise component is then: 

\begin{align}
k(\tau) = S_0\omega_0 Q e^{-\frac{\omega_0 \tau}{2Q}}(\cos(\eta\omega_0 \tau) + \frac{1}{2\eta Q}\sin(\eta \omega_0 \tau)),
\end{align}
where $\eta = |1-(4Q^2)^{-1}|^{1/2}$. Following the recommendation of
\cite{celerite}, we combine two of these kernels to model the stellar
variations.
In the first element of our combined kernel, $Q$ is fixed to $\frac{1}{\sqrt{2}}$ while $S_0$ and $\omega_0$ are allowed to vary in our fits; this
represents a non-oscillatory component that decays rapidly with
$\tau$.  In the second component, all three parameters are allowed to vary in order to model the
oscillatory component of the stellar noise, which shows a dominant
frequency of approximately 21 d\textsuperscript{-1} in a Lomb-Scargle
periodogram.  We give the second $\omega_0$ an appropriate name: $\omega_{stellar}$. This is the parameter presented in Table \ref{table:all_results}.  Our noise model thus consists of 6 parameters: $S_0, \omega_0$ for the
first kernel; $S_0, Q, \omega_0$ for the second kernel; and a white
noise term.

\subsection{Markov Chain Monte Carlo Fits}
We explore the parameter space for our model and determine best-fit
parameters using a Markov chain Monte Carlo (MCMC) analysis.  We carry
out our fits using the \emcee package \citep{emcee}, which is a Python
implementation of an affine-invariant ensemble sampler.
This approach allows for more efficient exploration of highly correlated parameters spaces, as
proposed steps are generated using an ensemble of walkers whose
positions are distributed along the regions of highest probability.
Our models have $24-48$ free parameters, depending on planet and
wavelength, so we carry out our fits using 250 walkers in order to ensure sufficient sampling of the parameter space. 

We obtain a starting point for our fits by using the published transit
and eclipse parameters to get a model light curve, dividing the data
by the model, and obtaining PLD parameters by fitting the residuals using linear regression.  For WASP-33, the
Gaussian process parameters were estimated by plotting the
autocorrelation and manually tweaking the parameters until we achieved a good
match.  We then generated initial positions for each of the 250 walkers by taking the reference values calculated above and
randomly perturbing each dimension.  Each dimension is first perturbed by a number drawn from a normal
distribution with a mean of 0 and standard deviation of 1\% of the
nominal value.  We then do an absolute perturbation, with each
dimension being perturbed by a number drawn
from a normal distribution with a mean of 0 and standard deviation of
0.01.  This two-step perturbation ensures that dimensions whose
initial values
are 0, as well as dimensions whose initial values are far
from 0, are both sufficiently spread out.  Although this is a very broad distribution relative to the final uncertainties in these parameters, it ensures that our fits are able to reliably identify the global maximum in the likelihood function.

We first run \emcee for 100,000 steps, resulting in a total of 25
million steps in the combined chain.  We then take the single step
with the highest likelihood from this chain, which should be very
close to the global maximum, and initialize a new set of 250 walkers
to a Gaussian ball centered around that point. The standard deviation
of this ball for a given dimension is $10^{-4}$ of the initial value.
We then burn in this new set of walkers and run \emcee for an additional 100,000
steps.  We determine our final posterior probability distributions
using the last half of this chain, and confirm that the acceptance rate for this section of the chain is between 20 and 50 percent.
We visually inspect the progress of randomly chosen walkers
to check that there is no overall trend, and that the number
of steps is appreciably larger than the period of quasiperiodic oscillations,
if any.  
Finally, we check for convergence by calculating the average autocorrelation length for all
walkers and ensure that the number of steps for each walker is at
least ten times the length for each model parameter.  

\subsection{Model Selection and Optimization of Photometry}
For each planet and each wavelength, we need to choose the optimal
order ($n$ in Equation \ref{eq:pld}) for our PLD model, the photometric aperture used to generate our light curve, and the size of the bins used in the fits.  As discussed in \cite{pld_paper} and \cite{kammer_2015}, the PLD method performs better when it is fit using binned light curves.

We first determine the optimal PLD order by fitting light curves
generated using an aperture of 2.5 and a bin size of 128, which are
representative of the optimal values in previous fits to \spitzer data
sets (e.g., \citealt{wasp14b_phase_curve}).  We run MCMC fits for
models with PLD orders ranging between 1 and 5.  In each fit, we
calculate the Bayesian Information Criterion (BIC) value \citep{kass_1995} for every position in the chain,
and then calculate the median BIC over the entire chain.  We then select the order
with the lowest median BIC for our final version of the model.  A spot
check reveals that using maximum BIC instead of median BIC does not
change the result.

We next choose the optimal photometric aperture by repeating our MCMC fits to photometry generated using all 20 apertures, where we fix the order of our PLD model to the optimized value and keep the same bin size as in our previous fits.  In this case all of our models have the same number of free parameters, and we therefore select the aperture that produces the highest median
likelihood over our MCMC chain.  We also consider an alternative
aperture selection metric where we compare the median (best-fit) white
noise parameters for each aperture on the assumption that the best
aperture should have the smallest white noise value.  We find that
this gives very similar results to our previously adopted likelihood
metric.  We list our final choice of aperture for each observation in
Table \ref{table:observations}.

The bin size is more complicated to optimize, as it represents a trade-off between minimizing the noise on short versus long timescales.  
If it is too big, we average over pointing variations and this can degrade the quality of the PLD model and increase the uncertainties in our model parameters.  If it is too small, the PLD parameters adjust themselves in such a way as to
minimize residuals on the shortest timescales ($\sim$seconds) at the expense of large timescale
residuals--even though the latter is closer to the timescale of the astrophysical variations and can bias our estimates for the astrophysical model parameters.  We consider bin sizes ranging from 1 to 4096 and find
that a bin size of 128, corresponding to a time interval of 51
seconds, is a good
compromise.  51 seconds is much longer than the shortest pointing
jitter variations, which have a timescale of seconds, but much shorter
than the timescale of any astrophysical signal.  We therefore use 128 as the bin size for our final fits
to all four phase curves.

The optimal results are presented in Table \ref{table:observations}.

\section{Discussion}
\label{sec:discussion}

\begin{figure*}[t]
  \includegraphics [width= 0.5\textwidth]{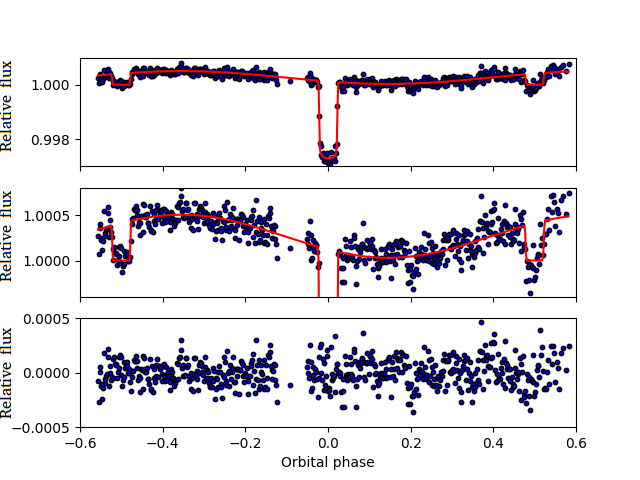}
  \includegraphics [width= 0.5\textwidth]{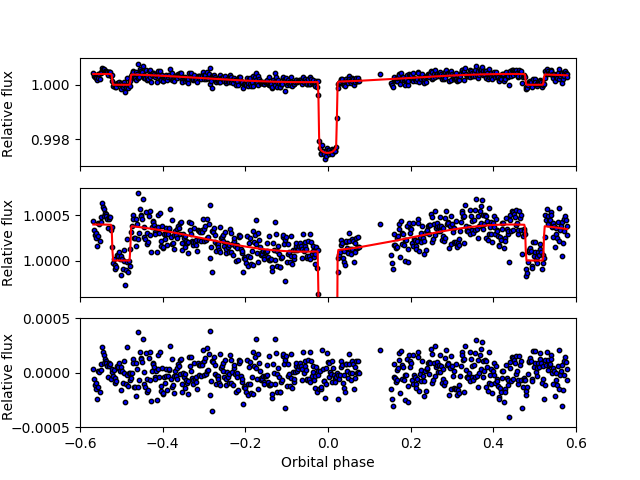}
  \caption{Normalized light curve for our 3.6 \um (left) and 4.5 \um (right)
    observations of HD
    149026b with our instrumental noise model divided out (blue filled
  circles) and a representative model fit overplotted for comparison
  (red line).  The upper two panels show the same light curve and
  model with a different $y$ axis range.  We show the residuals from
  this solution in the lower panel.  All three panels use a bin size
  of 1024 points (6.8 minutes).}
\label{fig:hd149026_fit_residuals}
\end{figure*}

\begin{figure*}[t]
  \includegraphics [width= 0.5\textwidth]{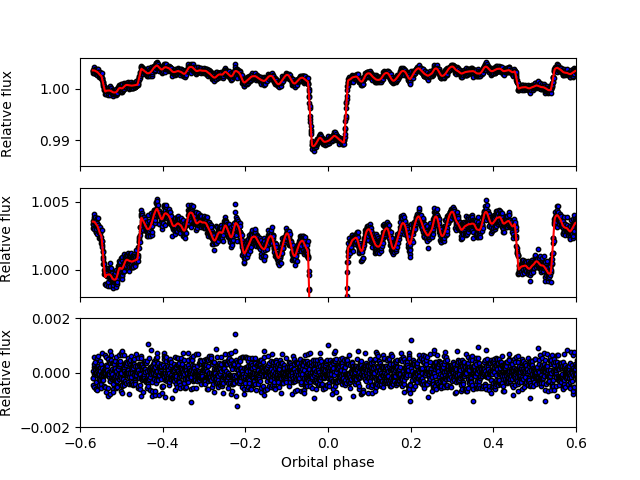}
  \includegraphics [width= 0.5\textwidth]{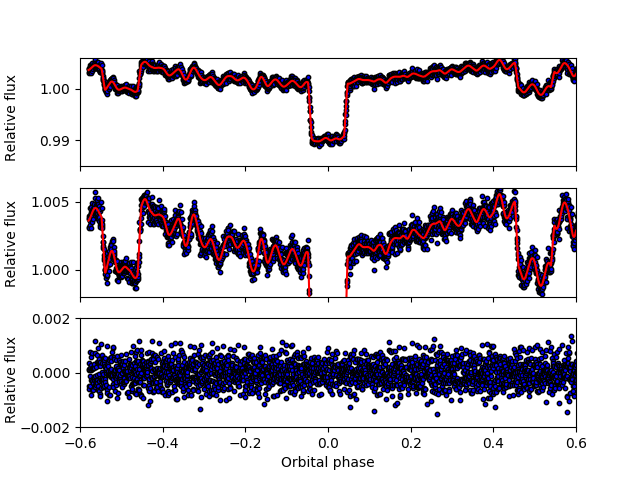}
\caption{Normalized light curve for our 3.6 \um (left) and 4.5 \um (right) observations of
  WASP-33b with the instrumental noise model divided out (blue filled
  circles) and a representative model, including the Gaussian Process stellar pulsation model,
  overplotted for comparison
  (red line).  All three panels use a bin size of 128 points (51
  seconds).  See Figure \ref{fig:hd149026_fit_residuals} caption
  for more details.}
\label{fig:wasp33_fit_residuals}
\end{figure*}

\begin{figure*}[h]
  \includegraphics [width=\textwidth]{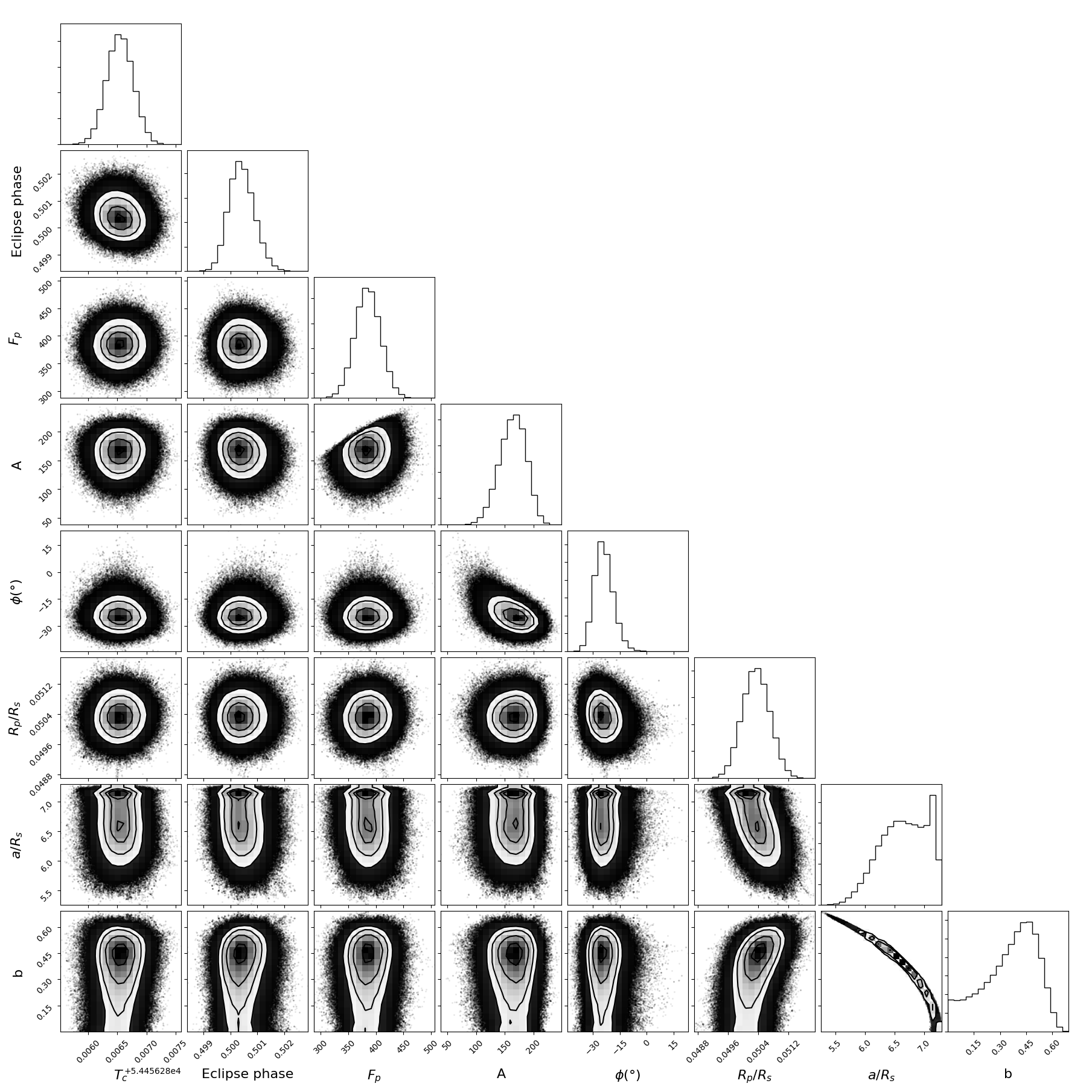}
  \caption{Posterior probability distributions for our fit to HD 149026b's 4.5 \um phase curve; also known as a triangle plot.  The triangle plot for HD 149026b 3.6 \um is similar, while those for WASP-33b are more Gaussian and less correlated.}
  \label{fig:hd149026_ch2_triangle}
\end{figure*}

Figures \ref{fig:hd149026_fit_residuals} and
\ref{fig:wasp33_fit_residuals} each show the phase curve model,
the systematics-corrected observations, and the fit residuals for the
highest likelihood iteration of the MCMC chain.  Figure \ref{fig:hd149026_ch2_triangle} shows the posterior probability distributions for our HD 149026b 4.5 \um model.  The
  triangle plot for our HD 149026b 3.6 \um model has similar correlations, while
the model parameters for WASP-33b are much more Gaussian and less correlated.

In Table \ref{table:all_results}, we present the results of our fits, including the secondary eclipse depth,
phase amplitude, and phase offset at 3.6 and 4.5 \um for each planet.
Here, the secondary eclipse depth ($F_p$) is defined as the planetary
flux at the center of eclipse divided by the stellar flux.  Unlike the most
commonly used secondary eclipse model, which assumes that the
brightness from the planet is constant over the duration of the
eclipse, this model allows for variations in the planet's brightness
during this interval (see \citealt{lewis_2013} for a discussion of the importance of this approach when fitting planets on highly eccentric orbits).  The phase amplitude is
defined as the amplitude of the sinusoidal phase curve, $A = \sqrt{c_1^2 + c_2^2}$. 
The phase offset is the difference in degrees between the secondary
eclipse center and the phase curve maximum, with a negative offset meaning the maximum occurs before the center of the
eclipse.

In Table \ref{table:all_results}, we also present noise properties for each light curve.  These include
the measured white noise ($\sigma_{\rm white}$), the theoretical photon noise
($\sigma_{\rm photon}$), the standard deviation of the residuals of the best fit model ($\sigma_{\rm tot}$), and the lag-1 autocorrelation of said residuals--the last being a measure of correlated noise.

\begin{table*}
  \centering
  \caption{Best-Fit Parameters}
  \begin{tabular}{C C C C C}
    \hline
    \text{Parameter} & \text{WASP-33b 3.6 \um} & \text{WASP-33b 4.5 \um} &
      \text{HD 149026b 3.6 \um}\footnote{The HD 149026b phase
        amplitude and offset should be treated with skepticism due to
        data quality issues; see Subsection \ref{subsec:hd149026b_ch1}} & \text{HD 149026b 4.5 \um}\\
      \hline
      \text{Eclipse depth (ppm)} & 3506 \pm 173 & 4250 \pm 160 & 430 \pm 19 & 385 \pm 23\\
      \text{Amplitude (ppm)} &  936 \pm 105 & 1792 \pm 94 & 189^{+27}_{-39} & 164^{+22}_{-24}\\
      \text{Phase offset } (\degree) & -12.8 \pm 5.8 & -19.8 \pm 3.0 & 32.2^{+17}_{-15} & -24.3^{+5.5}_{-4.7}\\
      \text{Transit center } (\text{BJD\textsubscript{UTC}}) & 2456029.62604 \pm 0.00016 & 2456024.74659 \pm 0.00014 & 2455661.78488 \pm 0.00021 & 2455673.28848 \pm 0.00022\\
      R_p/R_s & 0.108 \pm 0.001 & 0.103 \pm 0.0011 & 0.0519 \pm 0.0004 & 0.0503 \pm 0.0004\\
      a/R_* & 3.65^{+0.03}_{-0.05} & 3.65^{+0.04}_{-0.05} & 6.38^{+0.5}_{-0.4} & 6.67 \pm 0.4\\
      b & 0.150^{+0.072}_{-0.089} & 0.16^{+0.08}_{-0.10} & 0.48^{+0.09}_{-0.15} & 0.38^{+0.12}_{-0.20}\\
      i & 87.6^{+1.4}_{-1.2} & 87.6^{+1.5}_{-1.3} & 85.6^{+1.6}_{-1.1}
      & 86.7^{+1.7}_{-1.4}\\
      \phi_{\rm eclipse} & 0.50023 \pm 0.00028 & 0.50045 \pm 0.00024 & 0.4989 \pm 0.00033 & 0.50039^{+0.00050}_{-0.00044}\\
      \text{slope (ppm/day)} & 1590 \pm 203 & 60 \pm 170 & 510 \pm 130, 140 \pm 100 & -81 \pm 28\\
      \sigma_{\rm white} \text{(ppm)} & 356 \pm 5.6 & 451 \pm 7 & 270 \pm 4.4, 305 \pm 4.2 & 348 \pm
      3.6\\
      \sigma_{\rm photon} \text{(ppm)} & 306 & 411 & 227 & 313\\
      \sigma_{\rm tot} \text{(ppm)} & 628\footnote{After subtracting the Gaussian Process stellar pulsation model, the standard deviation decreases to 343 ppm for the best-fit model.} & 708\footnote{Post-GP: 438 ppm} & 290 & 347\\
      \omega_{\rm stellar} \text{ (rad/day)} & 129 \pm 1.3 &
      130 \pm 1.5 & - & -\\
      \text{Lag-1 autocorrelation} & 0.69\footnote{Post-GP: -0.027} & 0.60\footnote{Post-GP: -0.047} & 0.08 & -0.02\\
      \hline
  \end{tabular}
  \label{table:all_results}
\end{table*}

\subsection{Overall quality of fits and problems with the HD 149026b 3.6 \um phase curve}  
\label{subsec:hd149026b_ch1}
For every observation except HD 149026b at 3.6 \um, our higher order PLD model appears to provide a satisfactory fit to the data; as shown in Table \ref{table:all_results}, these observations have a measured white noise only 10-20\%
higher than the photon noise limit.  As shown in Figure
\ref{fig:hd149026_fit_residuals}, the HD 149026b 4.5 \um
observations have reasonable residuals, with no prominent unremoved systematics.  WASP-33b residuals are harder to evaluate visually, but the small measured white noise indicates that most sources of error other than photon noise have been accounted for.

In contrast to the good general quality of the other light curves, the HD 149026b 3.6 \um observation should be treated with
skepticism.  We find that the data strongly prefer a large positive phase offset, which
is inconsistent with the negative offset at 4.5 \um and is difficult
to reproduce with thermal emission from standard GCMs assuming
synchronous rotation \citep{heng_showman}.

The data themselves are of unusually low quality. 
This observation is divided into two segments with a 2.4 hour gap, corresponding to a telescope downlink break. 
In each segment the star's position varies over an approximately oval region with a dimension of 0.5 pixels in the $x$ direction and 0.2 pixels
in the $y$ direction, and the two oval regions are themselves separated
by 0.5 pixels.  During the first segment the star is relatively close
to the center of the central pixel, but after the repointing required
for the data downlink the star's position in the second segment falls
on an adjacent pixel.  It is very close to the edge, with that pixel
receiving 40\% of the light and the second
brightest pixel receiving 20\%.

All of this bodes poorly for PLD correction, or for any other
correction algorithm.  Not surprisingly, we find that the RMS of the fit residuals for the second segment is 17\% higher than in the first segment, providing tangible evidence for the persistence of these edge effects.  It should be noted that none of these problems
appear in the other three data sets.  Both WASP-33 observations are continuous, and the
star's position shifts over an area no bigger than 0.2 by 0.2 pixels.
Although the 4.5 \um observations for HD 149026
also include a downlink break in the middle, the telescope was able to
return the star to approximately the same position at the end of the
downlink and as a result the data from both segments span a single 0.7
x 0.2 pixel oval centered near the middle of the pixel.

We experimented with many different models for the data.  We initially
tried fitting the 3.6 \um HD 149026 data with a single systematics model utilizing the
same $3\times3$ pixel postage stamp centered on the middle of the array that we used for our other data sets.  
However, because the star is offset relative to this postage stamp during the second
segment of data, this fit resulted in prominent systematics in the residuals
for the second segment and a best-fit phase offset of approximately
60$\degree$.  We then considered a separate Gaussian Process noise
model for the two segments, where the second segment was represented
by a simple harmonic oscillator kernel.  The results did not change.
Other models we tried including fitting only the first segment and
introducing a separate linear slope for both segments.  

In the end, we settled upon a separate noise and systematics
model for each segment.  Each segment therefore has its own PLD
parameters, error parameter, and linear slope.  This drastically reduced the
residual systematics in the second segment.  Compared to the model
where both segments had the same noise and systematics model, this
segmented model has 9 additional free parameters and $\Delta BIC = -926$.
We found a best-fit phase offset of approximately 30$\degree$,
compared to 60$\degree$ with the simpler model.

We stress that the phase offset and amplitude are likely unreliable
even in this improved version of the fits.  We adjusted the bin size to see what effect it has
on the phase offset, and found that it monotonically decreases from 80 degrees west to
80 degrees east as the bin size increases from 1 to 4096.  Similarly,
the phase amplitude ranges from 190 ppm to 950 ppm, although it does
not change monotonically with bin size.  Although we remain concerned about the reliability of the phase curve fit in this bandpass, we conclude that the measured transit should be relatively unaffected by these structures due to its short timescale and large amplitudes.  The secondary eclipse depth is somewhat more problematic.
As can be seen in Figure \ref{fig:hd149026_fit_residuals}, the light curve has visible systematics after the downlink break, with an upward fluctuation before the eclipse and a downward fluctuation during the eclipse.  These likely bias the estimated eclipse depth.  Despite this bias, we find that our 3.6 \um secondary eclipse depth is in agreement with the published value in this band from \citet{stevenson_2012}.

\subsection{Transit parameters and updated ephemerides}
We recovered the transit time, $a/R_*$, $i$, and $R_p/R_s$ from the
chain.  The results are shown in Table \ref{table:all_results}.
Notably, we do not see any of the transit anomalies seen by
\cite{kovacs_2013} in the WASP-33b light curves.  These
anomalies included a 8 mmag rise in brightness across the transit
(Figure 8 of \citealt{kovacs_2013}) and a 1.5 mmag
mid-transit bump (their Figure 10), both of which were seen by multiple observers.

\begin{table}[h]
  \caption{Updated Ephemerides for both planets}
  \begin{tabular}{C C C C C}
    \hline
    \text{Parameter} & \text{HD 149026b} & \text{WASP-33b}\\
    \hline
    \text{Period (d)} & 2.87588874 & 1.21987089 \\
    T_0 (\text{BJD\textsubscript{UTC}}) & 2454456.78760 & 2454163.22367\\
    \text{Error in period (d)} & 5.9\e{-7} & 1.5\e{-7}\\
    \text{Error in $T_0$ (d)} & 0.00016 & 0.00022\\
    \hline
    
  \end{tabular}
  \label{table:period_epoch}
\end{table}

We combined our best-fit transit times for both planets with
previously published transit times to calculate
updated ephemerides.  For HD 149026b, we used \cite{charbonneau_2006}, \cite{winn_2008}, \cite{nutzman_2009}, \cite{knutson_2009}, \cite{hd149026b_transit_spectrum} and \cite{stevenson_2012}.

\begin{figure}[h]
\includegraphics [width= 0.5\textwidth]{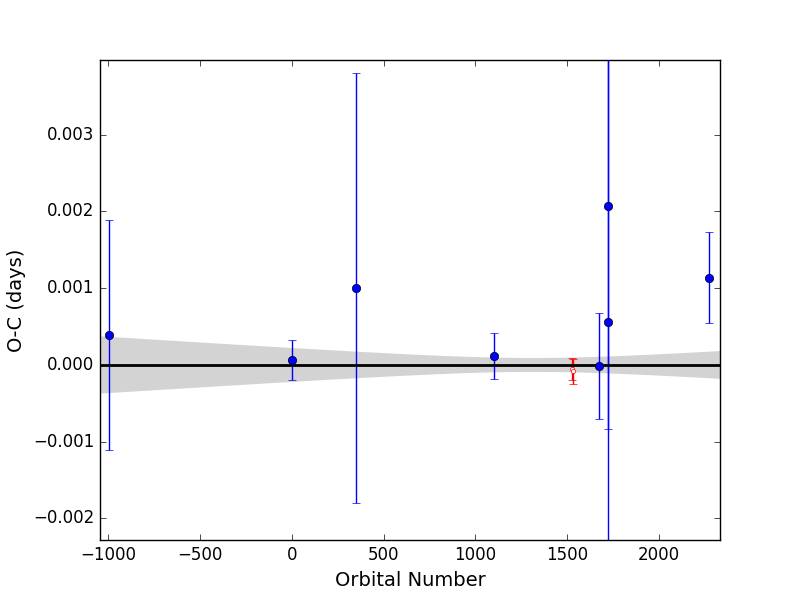}
\caption{Observed minus calculated transit times for WASP-33b calculated using our updated ephemeris.  Previously published results are shown as blue filled circles, and our results are shown as red open circles.  The black line indicates the predicted transit times at each epoch assuming a constant ephemeris, and the gray region indicates the 1$\sigma$ confidence interval.}
\label{fig:wasp33b_transits}
\end{figure}

\begin{figure}[h]
\includegraphics [width= 0.5\textwidth]{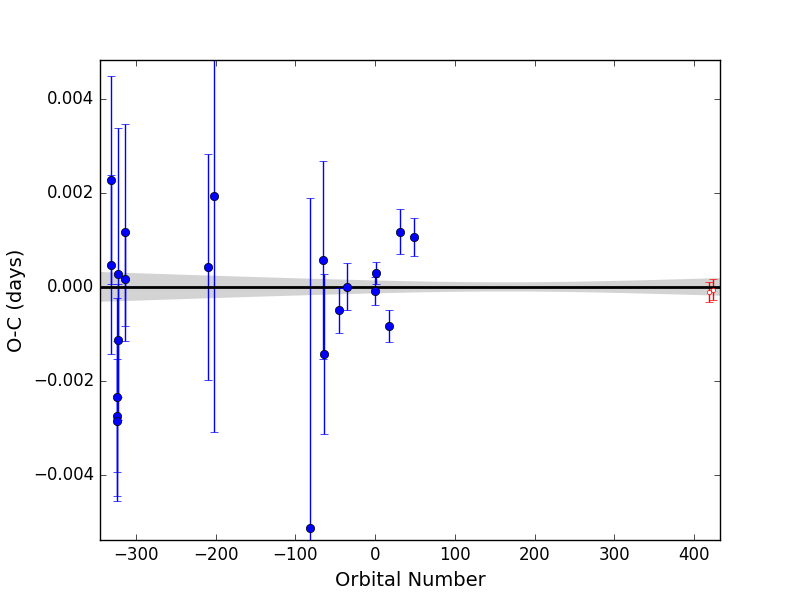}
\caption{Observed minus calculated transit times for HD 149026b calculated using our updated ephemeris; see Fig. 7 caption for more details.}
\label{fig:hd149026b_transits}
\end{figure}

For WASP-33b, we used \cite{wasp33b_discovery}, \cite{smith_2011},
\cite{kovacs_2013}, \cite{von_Essen_2014},
\cite{johnson_2015}, and \cite{turner_2016}.  The updated ephemerides are
shown in Table \ref{table:period_epoch}, and O-C plots for all
transits are shown in Figure \ref{fig:wasp33b_transits}
and \ref{fig:hd149026b_transits}.  We tested the goodness of fit with
$\chi^2$, finding that both planets are consistent with a linear
ephemeris--WASP-33b to well within 1$\sigma$, and HD-149026b to within
2$\sigma$ (p=0.07).  For HD 149026b, both the period and transit timing are fully
consistent with \cite{hd149026b_transit_spectrum}.  For WASP-33b, both
the period and transit timing are fully consistent with
\cite{kovacs_2013}.

It is notable that for both planets, the radius ratio is inconsistent
between the two channels, differing by 3\% for HD 149026b and 5\% for
WASP-33b.  This could be due to imperfect modeling of stellar
oscillations for WASP-33b, or to uncorrected systematics for both planets.
The difference corresponds to roughly five atmospheric scale heights for
HD 149026b and twelve scale heights for WASP-33b.  Similarly large discrepancies have
been reported in ground-based transit observations of other planets \citep[e.g.,][]{wasp36b_transit_spectrum}, but these appear to be inconsistent with most model predictions as well as space-based transmission spectroscopy of similar planets (e.g., \citealt{sing_2016}).

\cite{kovacs_2013} carried out a comprehensive analysis of
ground-based WASP-33b light curves, consisting of amateur and
professional data in the optical and near infrared bands.  They found
$R_p/R_s = 0.1143 \pm 0.0002$, which is a remarkable 6-10$\sigma$
higher than 
our \spitzer values.  However, the authors note anomalies in many of
their data sets, including a mid-transit hump, a skewed transit shape, and
discrepancies in transit depth measurements that are much larger than
the formal errors.

HD 149026b has more consistent transit depths in the literature. 
\cite{winn_2008} found $R_p/R_s = 0.0491^{+0.0018}_{-0.0005}$ in
Stromgren (b+y)/2 photometry, \cite{nutzman_2009} found $R_p/R_s =
0.05158 \pm 0.00077$ at 8 \um, and \cite{hd149026b_transit_spectrum}
found $R_p/R_s = 0.05416^{+0.00091}_{-0.00070}$ with NICMOS (1.1-2.0
\um).  The last measurement is $2-3\sigma$ higher
than our \spitzer values, but our results are
consistent with \cite{winn_2008} and \cite{nutzman_2009}.

For HD 149026b, our measured secondary eclipse depths are fully consistent with those
measured by \cite{stevenson_2012} in the same \emph{Spitzer} bands using BLISS mapping.  
For WASP-33b, the eclipse depth is in good agreement with
\cite{deming_2012} at 4.5 \um, although it is $1.7\sigma$ higher
  at 3.6 \um.  This might be because of imperfect modeling of stellar pulsations, leading to underestimated error bars in both papers.  Since we have a longer observational baseline over which to characterize stellar pulsations and measured two eclipses instead of one, our measurement of the eclipse depth should be less sensitive to the effects of stellar pulsations than \cite{deming_2012}.

\subsection{Constraints on atmospheric circulation}

\subsubsection{General circulation models}

We present cloud-free GCMs for HD 149026b and WASP-33b calculated using the
Substellar and Planetary Atmospheric Radiation and Circulation (SPARC)
model \citep{sparc_model}, which couples the MITgcm
\citep{adcroft_2004} to a two-stream implementation of the
multi-stream, plane-parallel radiative transfer code of
\cite{marley_1999}.  The MITgcm is an atmospheric and oceanic
circulation model that solves the primitive equations, which are
relevant for stably stratified atmospheres with large
horizontal/vertical aspect ratios (generally true for hot Jupiters).
The equations are solved using a finite-volume discretization on a
cubed-sphere grid, which allows longer time steps and increases the
accuracy near the poles as compared to a traditional
longitude-latitude grid.  The radiative transfer code employs the
correlated-k method with 11 bands optimized for accuracy and
computational efficiency. The opacities are calculated assuming local
thermodynamic and chemical equilibrium. This code has been used
extensively to model the atmospheric circulation of exoplanets over a
wide range of planetary properties (e.g.,
\citealt{lewis_2010,kataria_2015,kataria_2016,wakeford_2017}).  After
running the GCM, we extract light curves following the method of \cite{fortney_2006b}.

We list the predicted eclipse depths, amplitudes, and phase offsets at 3.6 and 4.5 \um for each model in Table \ref{table:gcm_params}; these can be compared directly to the measured values in Table \ref{table:all_results}.

\begin{figure}[h]
  \centering \subfigure[Solar metallicity] {\includegraphics
    [width=0.5\textwidth]{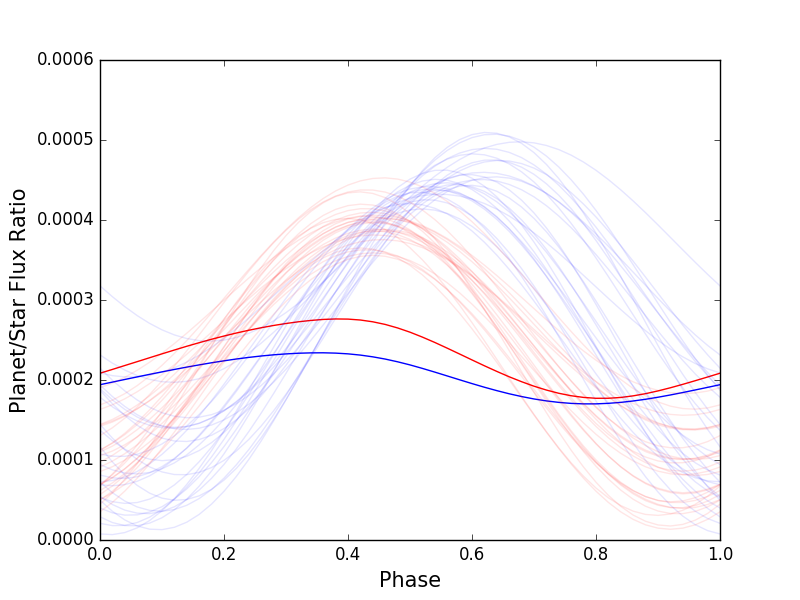}}\qquad
  \subfigure[$30\times$ solar metallicity] {\includegraphics
    [width=0.5\textwidth]{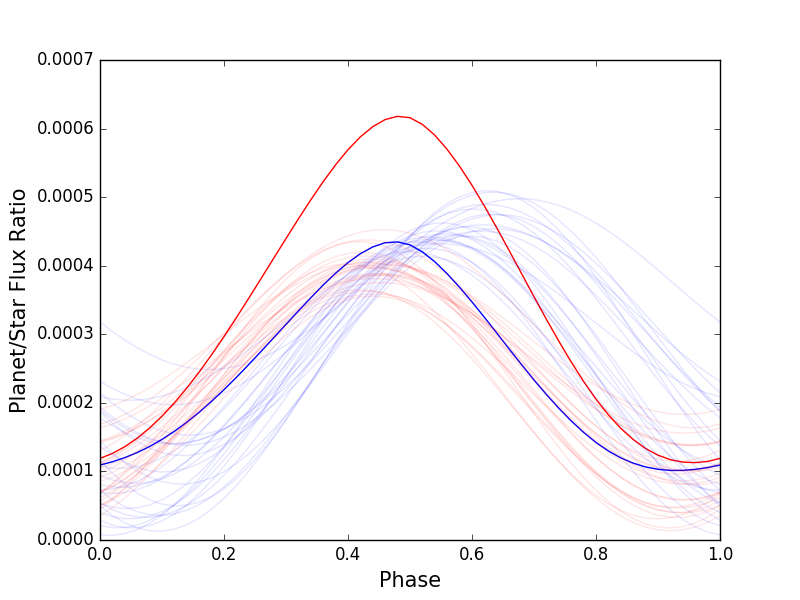}}
\caption{Comparison of the GCM-generated phase curves (thick lines) for HD 149026b with our measured phase curves (thin lines).  The 24 thin lines each represent one randomly selected MCMC
  step, and the dispersion in these lines is therefore representative of the uncertainties in the measured phase curve shape. No TiO is included.  3.6 \um results are plotted as blue curves while 4.5 \um results are in red.}
\label{fig:hd149026_gcm_comp}
\end{figure}

\begin{table*}
  \centering
  \caption{Phase curve parameters from GCMs}
  \begin{tabular}{c c c c C c C}
      \hline
      Planet & TiO? & Metallicity & Band
        (\um) & F_p \text{(ppm)} & A (ppm) & \phi (\degree)\\
      \hline
      WASP-33b & No & $1\times$ & 3.6 & 4086 & 1664 & -9.9\\
      WASP-33b & No & $1\times$ & 4.5 & 4597 & 1903 & -9.6\\
      WASP-33b & Yes & $1\times$ & 3.6 & 4151 & 1688 & -8.8\\
      WASP-33b & Yes & $1\times$ & 4.5 & 4779 & 1981 & -7.6\\ 
      HD 149026b & No & $30\times$ & 3.6 & 430 & 167 & -11\\
      HD 149026b & No & $30\times$ & 4.5 & 616 & 252 & -7\\
      HD 149026b & No & $1\times$ & 3.6 & 219 & 31.9 & -55\\
      HD 149026b & No & $1\times$ & 4.5 & 260 & 49.6 & -45\\
      \hline
  \end{tabular}
  \label{table:gcm_params}
\end{table*}

For HD 149026b we consider models with solar and $30\times$ metallicity and
compare the resulting phase curves to our best-fit phase curve
model in Figure \ref{fig:hd149026_gcm_comp}.  We only consider
models without TiO, as \cite{stevenson_2012} found that this planet's
dayside emission spectrum was best described by a model without a
temperature inversion.  We find that the solar metallicity GCM
predicts a relatively small phase curve amplitude in both bandpasses,
in sharp disagreement with our data.  The $30\times$ solar metallicity model
has a higher opacity in both \spitzer bandpasses and therefore probes
lower pressures (higher altitudes) than the solar metallicity model,
leading to larger predicted phase curve amplitudes.  This model comes
closer to matching the data, although it underestimates the amplitude
at 3.6 \um and overestimates it at 4.5 \um.  We note that neither the
1D models shown in \cite{stevenson_2012} nor the 3D GCMs are
able to match the measured secondary eclipse depths at 3.6 and 4.5
\um, and speculate that these discrepancies in both secondary eclipse
depths and phase curve amplitudes might be resolved by increasing the
amount of CO and/or CO$_2$ in the atmosphere.   Both CO and CO$_2$
have absorption bands in the 4.5 \um \spitzer bandpass, increasing
their abundance will accordingly decrease the planet's brightness in
this band relative to the 3.6 \um band.   Previous models for GJ 436b
\citep{moses_2013,morley_2017} serve as a useful demonstration of the
effect of very high atmospheric metallicities ($>200-300\times$ solar)
on the strength of the CO absorption in the 4.5 \um band.   On the planet's night side, which is cool enough to fall near the transition from CO to methane-dominated carbon chemistry, disequilibrium chemistry due to quenching and horizontal transport could increase the relative amount of CO and CO$_2$ \citep{cooper_showman_2006}. 

\begin{figure}[h]
  \centering \subfigure[Without TiO] {\includegraphics
    [width=0.5\textwidth]{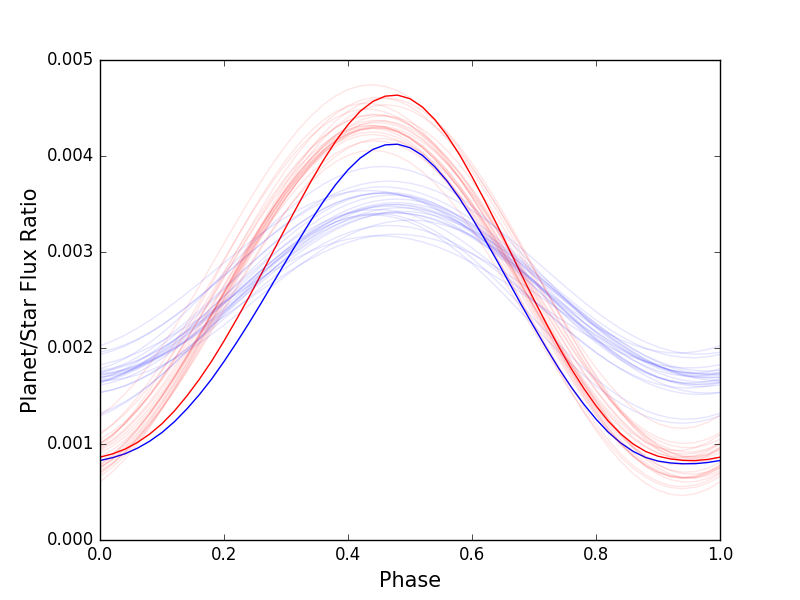}}\qquad
  \subfigure[With TiO] {\includegraphics
    [width=0.5\textwidth]{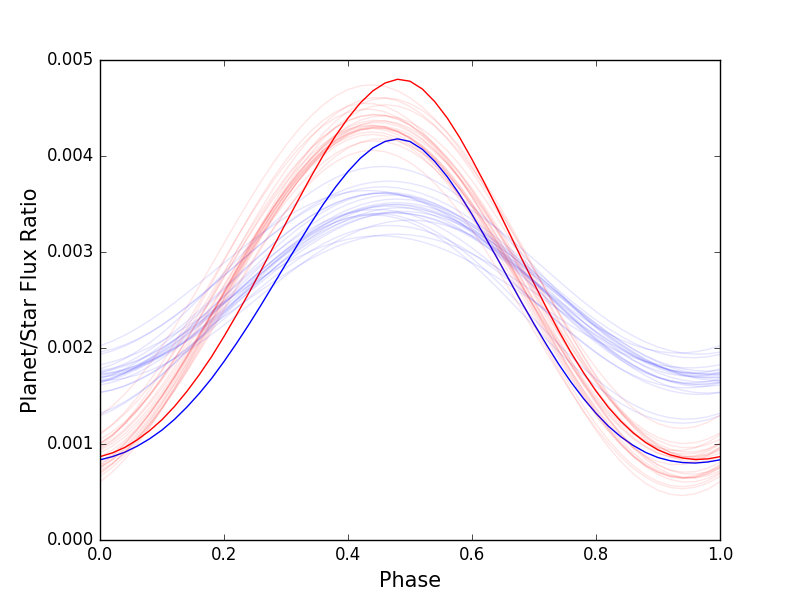}}
\caption{Comparison of GCM-generated phase curves (thick lines) for WASP-33b
  with our measured phase curves (thin lines).  The 24 thin lines each represent a randomly selected MCMC
  step, and the dispersion in these lines is therefore representative of the uncertainties in the measured phase curve shape. Both models assume solar
  metallicity.  3.6 \um results are plotted as blue curves and 4.5 \um results are in red.}
\label{fig:wasp33_gcm_comp}
\end{figure}

For WASP-33b, which has a much lower bulk density than HD 149026b, we
consider only the solar metallicity case for our GCMs.  As
discussed in Section \ref{sec:introduction}, this planet is one of the
most highly irradiated hot Jupiters discovered to date; its
dayside emission spectrum is best matched by models with a temperature
inversion and appears to hint at the presence of gas-phase TiO.  We
show predictions for two models in Figure \ref{fig:wasp33_gcm_comp}, including one with and the other without TiO, in order to evaluate the effect of this molecule on its dayside emission spectrum and day-night circulation.  We find that the differences in the phase curves for these two models are fairly subtle, and although the data are somewhat better matched by the model without TiO, both models disagree with the observations at the $2\sigma$ level.  In this case the models predict a larger phase curve amplitude and secondary eclipse depth in both bands.  

\begin{figure}[h]
\includegraphics [width= 0.5\textwidth]{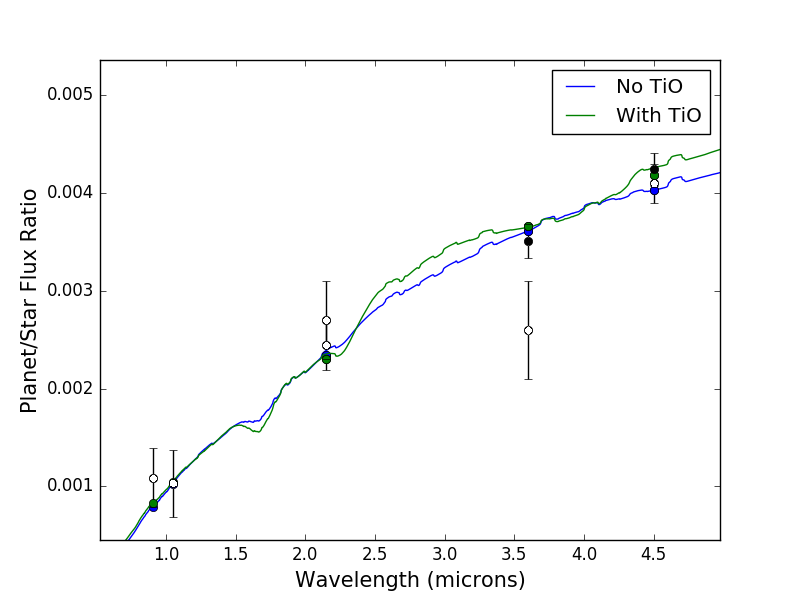}
\caption{Model emission spectrum for WASP-33b, compared with
  observations (black filled circles for our values, black open circles for literature values from \citealt{smith_2011,von_Essen_2015,deming_2012,de_Mooij_2013}).  Each blue or green point represents the band-averaged
  flux ratio corresponding to the observation at the same wavelength.}
\label{fig:wasp_spectrum}
\end{figure}

\begin{figure}[h]
\includegraphics [width= 0.5\textwidth]{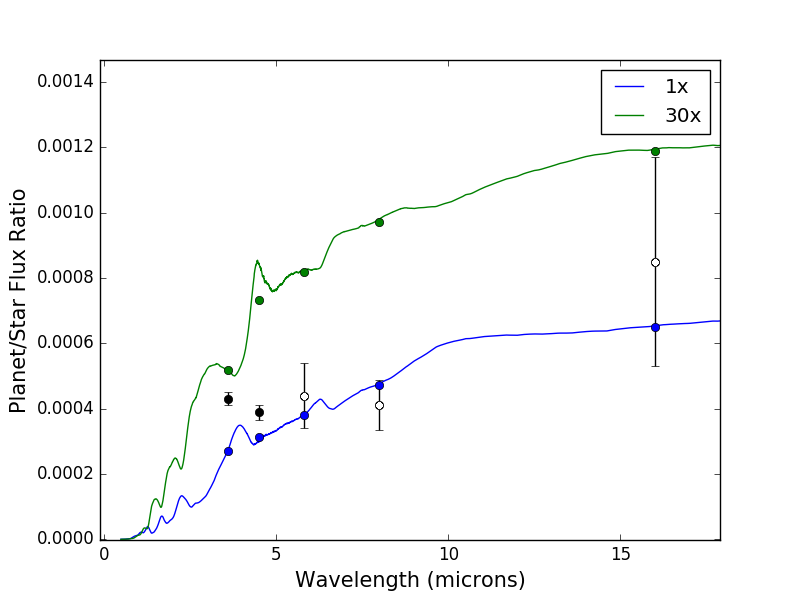}
\caption{Model emission spectrum for HD 149026b, compared with
  observations (black filled circles for our values, black open circles for literature values from \cite{stevenson_2012}). Each blue or green point represents the band-averaged
  flux ratio corresponding to the observation at the same wavelength.}
\label{fig:hd_spectrum}
\end{figure}

In Figure \ref{fig:wasp_spectrum} and \ref{fig:hd_spectrum}, we show
the GCM-derived emission spectra for the two planets.  Our WASP-33b
observations are consistent with both GCMs, which have very
similar emission spectra at \spitzer wavelengths.  HD 149026b, on the
other hand, is highly inconsistent with both GCMs in the \spitzer
3.6 and 4.5 \um bands.  Interestingly, eclipse observations in the 5.8, 8, and
16 \um \spitzer bands seem to favor the solar metallicity model over
the $30\times$ solar metallicity model.  We speculate that this could be because
the GCMs assume gas-only opacity with no hazes or clouds, resulting in a very low albedo; in contrast, our phase curve data appear to favor a high albedo
(see discussion in Subsection \ref{subsec:toy_model}).  A high albedo would cool the planet’s dayside, bringing the 30x model into bet- ter agreement with the data.  However, even with a reduced amplitude the 30x solar model is a poor match for the observed spectral slope across the 3.6-4.5 um bands; this could be explored with additional models in the future.  We note that our preference for the high metallicity model is primarily driven by the need to match the large observed phase curve amplitudes in both bands rather than the shape of the planet's dayside emission spectrum.  As an alternative to changing the planet's dayside albedo, the addition of localized clouds on the planet's night side could increase the phase curve amplitude for the solar metallicity model (e.g., \citealt{stevenson_2016}), bringing it into better agreement with our observations.

\subsubsection{A simple toy model for albedo and recirculation efficiency}
\label{subsec:toy_model}

In this section, we use a simple toy model first presented in
\cite{cowan_agol_2011} to calculate average brightness temperatures, albedos, and
circulation efficiencies for the two planets under the assumption that their thermal emission is well-approximated by a blackbody and that these observations probe a similar range in pressures across all wavelengths and longitudes.  
In this model, a planet's thermal phase curve is described by two parameters: a Bond albedo ($A_B$) and a
heat redistribution efficiency ($\varepsilon$).  Planets absorb a
fraction $1-A_B$ of the stellar flux on their daysides, redistribute this energy to the nightside
with the stated efficiency, and then emit as a blackbody.  In the
$\varepsilon=1$ case, the entire planet has the same temperature, and
energy balance gives $T_p = \frac{T_0}{\sqrt{2}}(1-A_B)^{1/4}$
where $T_0=\frac{T_s}{\sqrt{a_*}}$ and $a_*=\frac{a}{R_s}$.  For the $\varepsilon=0$ case (i.e., no heat redistribution to the night side), the corresponding dayside temperature is given by $T_d = (2/3)^{1/4}(1-A_B)^{1/4}T_0$
and the nightside temperature is zero.  If we define $\varepsilon$ so that it
linearly interpolates between these two extremes we arrive at an analytic description for $\varepsilon$ and $A_B$ as a function of the effective blackbody temperatures for the planet's dayside ($T_d$) and nightside ($T_n$):

\begin{align}
\varepsilon = \frac{8}{5 + 3(T_d/T_n)^4}\\ A_B = 1 - \frac{5T_n^4 +
  3T_d^4}{2T_0^4}.
\end{align}
Here, and throughout the paper, we define the nightside and dayside to
mean an orbital phase of 0 and 0.5, respectively.  Previous studies have alternated between this definition and one in which day and night correspond to the hottest and coldest hemispheres on the planet.

The hemisphere-averaged planet brightness temperature is given by Equation 6 in \cite{cowan_agol_2011}, 
and depends on two things: the
brightness temperature of the star at the observed wavelength, and the
ratio $\psi(\lambda)$ between normalized planetary flux and transit
depth at the observed wavelength.  We first calculate the stellar
brightness temperature in each band using the closest model in the
BT-NextGen (AGSS2009) spectral grid, as provided by the Spanish
Virtual
Observatory\footnote{\url{http://svo2.cab.inta-csic.es/theory/newov2/index.php}}.
The spectral grid spacing is fine enough that choosing the adjacent
model changes the brightness temperature by less than 1\%.  As a
check on BT-NextGen, we also calculated brightness temperatures using Phoenix
models \citep{husser_2013}, and found that the results differed on
average by only 0.4\%.  We next calculate the planet-star brightness ratio $\psi(\lambda)$ for the day side by dividing the eclipse depth $F_p$ by the transit depth, and for the night side by
dividing $F_p - 2A\cos{\phi}$ by the transit depth, where $A$ is the
phase amplitude and $\phi$ is the phase offset.  We then convert this to a brightness temperature for the planet using the stellar brightness temperature calculated earlier.  

We obtain uncertainties on these brightness temperature estimates using the posterior probability distributions from our MCMC fit.  For each step in our MCMC chain,
we calculate $\psi(\lambda)$ for the day and night sides from the
chain itself.  With these parameters, we then calculated T\textsubscript{d} and T\textsubscript{n} in each bandpass, using the error-weighted average of both bandpasses to calculate the planet's corresponding albedo
and recirculation efficiency.  Finally, we compare to Figure 7 in \cite{cowan_agol_2011} by calculating the
quantities $T_{\varepsilon=0} = (2/3)^{1/4}T_0$ and
$T_d/T_0=(1-A_b)^{1/4}(\frac{2}{3} - \frac{5}{12}\varepsilon)$ for each planet.  
Although WASP-33b has a higher $T_{\varepsilon=0}$ than all of the planets in
Figure 7 of \cite{cowan_agol_2011}, we find that its temperature ratio is fully consistent with that of other highly irradiated
($T_{\varepsilon=0} > 2500 K$) planets.  Similarly, $T_d/T_0$ for HD 149026b is in good agreement with the values for other planets with similar irradiation levels despite lingering questions about the reliability of the 3.6 \um results.  We list the relevant values for each planet in Tables
\ref{table:toy_model} and \ref{table:toy_model_combined}.

\begin{table*}
  \centering
    \caption{Dayside and nightside brightness temperatures, Bond albedo,
    and recirculation efficiency for each channel}
  \begin{tabular}{c C C C C C}
      \hline \text{Planet} & \lambda(\mu m) & T_{b,day}(K) & T_{b,night}(K) & A_B & 
      \varepsilon \\ \hline \text{WASP-33b} & 3.6 & 3082 \pm 92 &
      1952^{+125}_{-134} & 0.25^{+0.09}_{-0.10} & 0.34 \pm
      0.06\\ \text{WASP-33b} & 4.5 & 3209^{+89}_{-87} &
      1498^{+114}_{-118} & 0.25^{+0.08}_{-0.09} & 0.12 \pm
      0.03\\ \text{HD 149026b} & 3.6 & 1941 \pm 46 & 1133^{+290}_{-270}
      & 0.36^{+0.10}_{-0.16} &
      0.26^{+0.26}_{-0.16}?\\ \text{HD 149026b} & 4.5 & 1649 \pm 49 &
      1018^{+115}_{-116} & 0.66^{+0.05}_{-0.06} &
      0.31^{+0.11}_{-0.10}\\ \hline
  \end{tabular}
  \label{table:toy_model}
\end{table*}

\begin{table}
  \centering
    \caption{Averaged brightness temperatures, Bond albedo,
    and recirculation efficiencies}
  \begin{tabular}{C C C}
      \hline
      \text{Parameter} & \text{WASP-33b} & \text{HD 149026b}\\
      \hline
      T_{\rm day} & 3144 \pm 114 & 1804 \pm 98\\
      T_{\rm night} & 1757 \pm 88 & 1032 \pm 120\\
      A_B & 0.25^{+0.09}_{-0.10} & 0.53^{+0.09}_{-0.11}\\
      \varepsilon & 0.22^{+0.05}_{-0.04} & 0.24^{+0.11}_{-0.09}\\
      T_{\varepsilon=0} & 3514 \pm 30 & 2276 \pm 37\\
      T_d/T_0 &  0.81 \pm 0.04 & 0.72 \pm 0.04\\
      \hline
  \end{tabular}
  \label{table:toy_model_combined}
\end{table}

\subsection{Comparison with other planets}
\label{subsec:comparison}
\begin{figure}[h]
  \centering \subfigure[All wavelengths included] {\includegraphics
    [width=0.5\textwidth]{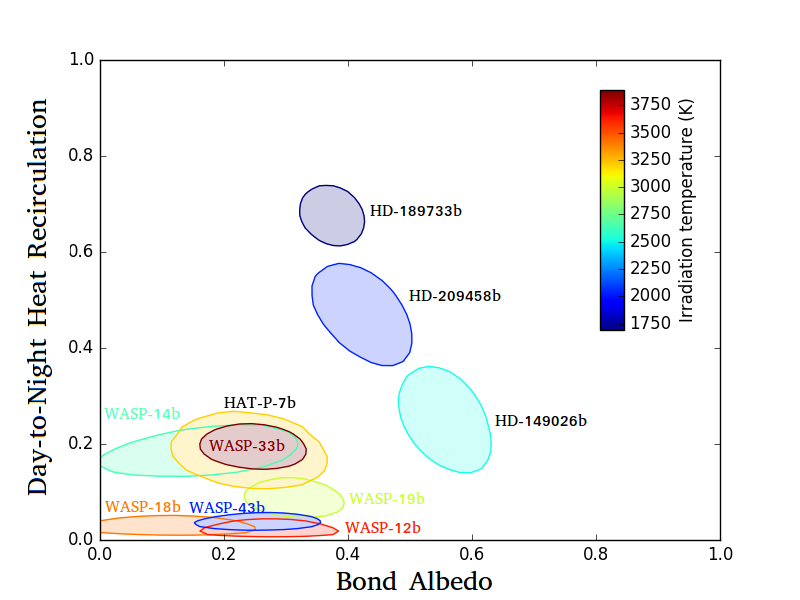} \label{subfig:albedo_eff_all_wavelengths}}\qquad \subfigure[Only \spitzer 3.6 and 4.5 \um data included] {\includegraphics
    [width=0.5\textwidth]{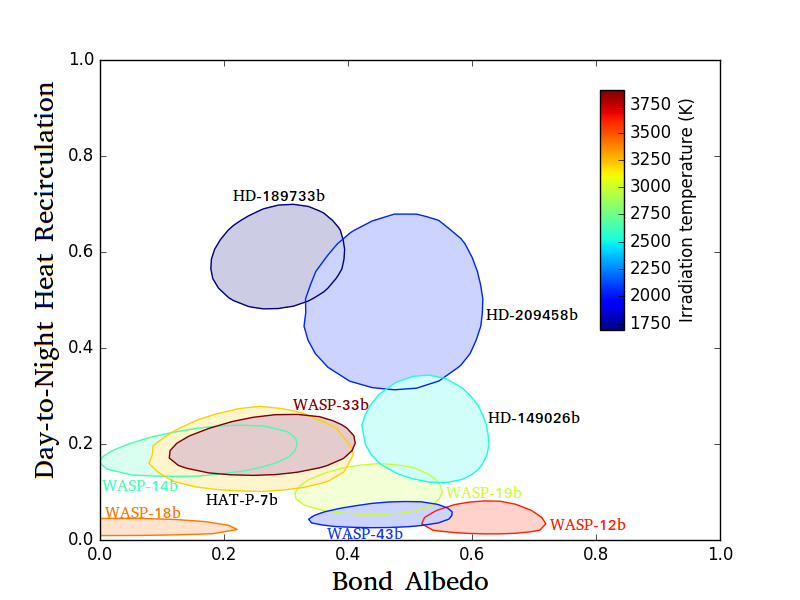}}
\caption{Albedo and recirculation efficiency for all exoplanets with published infrared phase curves, calculated following \cite{schwartz_2017} assuming the planet radiates as a blackbody.}
\label{fig:albedo_eff}
\end{figure}

WASP-33b is very unusual among the more than two hundred hot Jupiters
discovered to date, being the second most irradiated hot
Jupiter currently known (KELT-9b being the first).  Despite this
peculiarity, its albedo and recirculation efficiency appear largely
similar to those of other hot Jupiters observed to date.
In order to compare our planets to other hot Jupiters, we produce an updated version of Figure 3 from
\cite{schwartz_2017}, which plots contours corresponding to the albedo
and efficiency values estimated in \ref{subsec:toy_model}.  Our
version of the plot is shown in Figure \ref{subfig:albedo_eff_all_wavelengths}.  Although we largely follow the method described in this paper, our approach differs in several aspects:

\begin{enumerate}
\item Uncertainties are propagated using a Monte Carlo method, instead
  of dividing up the albedo-efficiency parameter space into cells and
  computing $\chi^2$ for each cell. 
\item If the dayside or nightside flux in an iteration is negative, we
  exclude the entire iteration, while \cite{schwartz_2017} set the
  corresponding temperature to zero.  This tends to slightly lower the recirculation
  efficiency.
\item \cite{schwartz_2017} assume a geometric albedo of 7\% and subtract the
  reflected light eclipse depth from the measured eclipse depth.  The
  actual geometric albedos of these planets are poorly constrained by current observations and an assumed albedo of 7\% has a negligible effect on our results, so we instead assume an albedo of zero.
  \end{enumerate}  
  Another complication is in the treatment of WASP-12b.  The WASP-12b phase curve paper \citep{wasp12b_phase_curve} included results from two
  analysis methods: polynomial fitting and point-by-point
  decorrelation.  We used the latter set of phase curve and eclipse
  depth parameters in our paper as it results in a more consistent phase curve offset between the two bands.  Although the two methods produced similar phase curve parameters for
  WASP-12b at 4.5 \um, they
  were very different for 3.6 \um, and our results differ
  substantially depending on which version we choose.  We downloaded
  the data in each bandpass ourselves, as well as an additional pair
  of phase curve observations taken in 2013.  All four data sets
  were analyzed with the
  same higher-order PLD approach used for WASP-33b and HD 149026b. For
  4.5 \um, our two results were consistent with each other and
  with both methods in \cite{wasp12b_phase_curve}.  For 3.6 \um, our
  two results were consistent neither with each other nor with either method
  in \cite{wasp12b_phase_curve}.  We therefore conclude that the
  properties of WASP-12b are not well-constrained by the current observations, although we still show it in our plots.

 \begin{figure}[h]
\includegraphics [width= 0.5\textwidth]{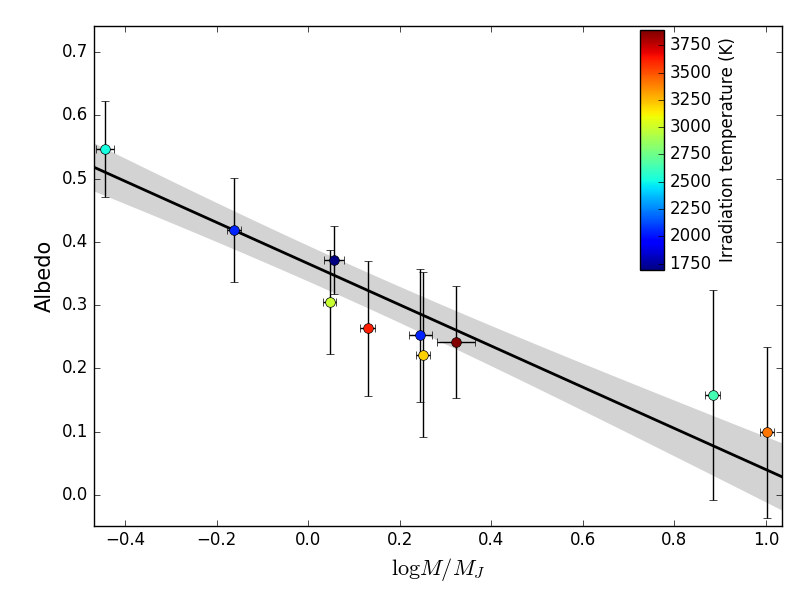}
\caption{Logarithm of mass vs. albedo for exoplanets with phase
  curves.  Albedo is calculated using the simple toy model described in \cite{schwartz_2017}, in which the planet is assumed to radiate as a blackbody.  We also overplot the best-fit linear function as a black line, with the 1$\sigma$ confidence interval shown in gray.}  
\label{fig:mass_vs_albedo}
\end{figure}

We show two versions of the albedo-efficiency plot in Figure \ref{fig:albedo_eff} ,
including one with all published thermal secondary eclipse and phase
curve data and another which only considers 3.6 and 4.5 \um \spitzer
data in order to ensure a more uniform analysis.  The only planet that
moved significantly was WASP-12b.

This figure shows that WASP-33b has a Bond albedo and
recirculation efficiency that appear largely similar to those of other hot Jupiters despite its high irradiation level.  HD 149026b, however, appears to have an unusually high
albedo in our toy model.  Its best-fit albedo is
higher than that of all other planets with thermal phase curves, and it is
also higher than any of the optical geometric
albedos measured by Kepler as shown by Figure 7 of
\cite{schwartz_cowan}.  This might reasonably be explained by the presence of a reflective cloud layer in this planet's upper atmosphere; the presence of such a cloud layer could be confirmed with future transmission spectroscopy.

\cite{parmentier_2016} calculated the effective cloud coverage for a range of equilibrium temperatures, three cloud top pressures, and a number of cloud compositions.  Although they focused on investigating the role of clouds at optical wavelengths rather than in the \spitzer bands, we can nonetheless utilize their results to explore the potential cloud species that might be present in HD 149026b's atmosphere.  They report evidence for the presence of silicate clouds for $T_{eq} > 1600 K$, presence of MnS clouds for $T_{eq} < 1600 K$, and absence of silicate clouds for $T_{eq} < 1600 K$.  HD 149026b has a zero-albedo equilibrium temperature of 1700 K, very close to the 1600 K dividing line.  If there are silicate clouds, their Figure 13 shows that the dayside effective cloud coverage is expected to be 30-80\%, depending on the cloud top pressure, while the nightside coverage is 50-100\%.  If there are MnS clouds but no silicate clouds, the dayside cloud coverage would be 0-20\%, while the nightside cloud coverage would be 20-100\%.

\cite{mahapatra_2017} used a kinetic, non-equilibrium cloud formation model to study cloud structures and compositions.  For HD 149026b, they found that clouds are likely composed of many different species, with TiO\textsubscript{2} dominant at the cloud top ($10^{2.5}$ bar) while species such as Fe, SiO, and MgSiO\textsubscript{3} are common deeper down.

One of the most interesting aspects of Figure \ref{fig:albedo_eff} is
that it shows no obvious correlations between irradiation temperature,
albedo, and efficiency.  In fact, it appears that planets with very
different irradiation temperatures can have similar albedos and recirculation
efficiencies.  We plotted mass versus efficiency, irradiation
temperature versus albedo, and irradiation temperature versus
efficiency, finding that the two least massive
planets in our sample--HD 149026b and HD 209458b--are outliers in both albedo and
efficiency.  In \cite{wasp14b_phase_curve} we previously suggested a possible correlation between mass
and albedo, but after additional data was collected, we concluded in \cite{wong_2016} that a simple mass-albedo correlation was no longer
tenable.  However, we see a strong correlation when we plot these two parameters in Figure
\ref{fig:mass_vs_albedo}.  The linear model has a lower BIC than the
constant-albedo model ($\Delta BIC = -12.5$), indicating a strong
preference for the linear model.  In addition, the error bars on the
albedo seem overestimated, possibly due to the large systematic error
we deliberately introduce (in accordance with
\citealt{cowan_agol_2011}) in converting from brightness temperature to physical temperature.  After subtracting the best fit linear model, we
find $\chi^2 = 1.86$ for the residuals; with 8 degrees of freedom, there is only a 1.5\%
probability of obtaining a $\chi^2$ this low.  If the errors were
correctly estimated, $\Delta BIC$ would be even more negative,
preferring the linear model even more strongly.  Our best fit line has
slope $m=-0.326 \pm 0.047$ and intercept $b=0.366 \pm 0.016$.
Coincidentally, 0.366 is very close to the Bond albedo of Jupiter
itself.

The physical explanation for the decrease in albedo with mass is
unclear.  One possibility is that increased surface gravity makes it
harder for cloud particles to be kept aloft, as shown in Equation 10
of \cite{heng_2013}.  The increased cloudiness at low surface gravity
has been observed on brown dwarfs \citep{faherty_2016}.  The main
difficulty with this explanation is that we have also plotted the relation
between surface gravity and albedo, and although an anticorrelation is
seen, it is much less statistically significant than the mass-albedo
correlation ($\Delta BIC = -2.4$, compared to $\Delta BIC = -12.5$).

\subsubsection{Phase curve offsets}

The toy model discussed above derives recirculation efficiency from
the observed nightside flux.  However, the nightside flux can only be
measured by reference to the secondary eclipse, which is hours or days
away.  This makes the measurement
particularly sensitive to instrumental noise sources on long
timescales, including the long-term pointing drift present in many
phase curve observations.  As an example, the \spitzer phase curves for
WASP-43b imply a negative nightside flux, which is
unphysical \citep{keating_2017}.

In this section we explore correlations between phase offset and other
planetary parameters.  In GCMs, both the size of the phase
offset and the relative temperature gradient between the day and night
sides increase with increasing depth (pressure) in the atmosphere
(e.g., \citealt{sparc_model}).  We therefore consider whether or not the measured phase offset might be useful as a proxy for
recirculation efficiency.  We plotted recirculation efficiency against
phase offset and found that although planets with very large phase offsets have somewhat high
efficiencies and planets with very small phase offsets have somewhat low
efficiencies, the correlation is by no means exact.  We conclude that either phase offset
is an imperfect proxy for recirculation efficiency in practice, or the
recirculation efficiency calculated using our simple toy model is simply not accurate enough for
the correlation to be obvious.

\begin{figure}[h]
\includegraphics [width= 0.5\textwidth]{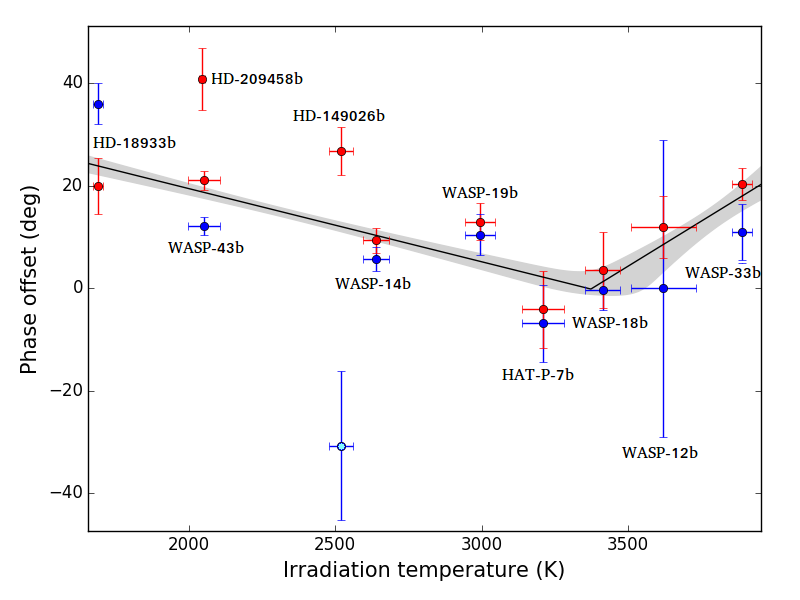}
\caption{Phase offset vs. irradiation temperature for all hot Jupiters
  on circular orbits with thermal phase curves.  Blue represents 3.6
  \um while red represents 4.5 \um.  The light blue point
  is the 3.6 \um observation for HD 149026b, which we discuss in
  \S\ref{subsec:hd149026b_ch1}. The black lines represent the best-fit
  bilinear model, while the gray
  region indicates the 1$\sigma$ confidence interval.}
\label{fig:T_vs_offset}
\end{figure}

\begin{figure}[h]
\includegraphics [width= 0.5\textwidth]{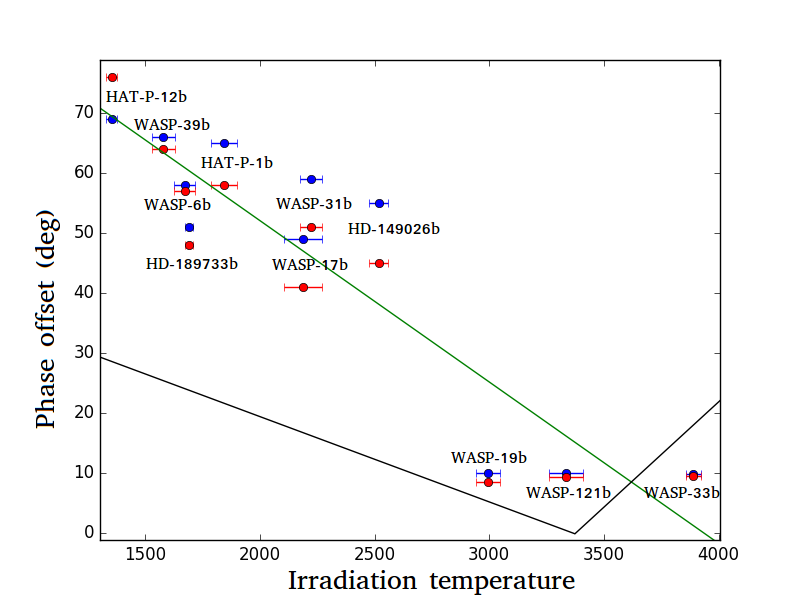}
\caption{Phase offset vs. irradiation temperature for GCM-modeled
  planets.  The green line is the best fit linear model to the GCM
  data, while the black lines represent the best fit bilinear model to the
  observations.  The black lines are identical to the ones in Figure \ref{fig:T_vs_offset}.}
\label{fig:model_T_vs_offset}
\end{figure}

Figure \ref{fig:T_vs_offset} shows a strong correlation between a
planet's irradiation temperature $T_0 = T_{\rm eff}/\sqrt{a_*}$ and its phase
offset.  There is a clear downward trend until $3400 K$, after which
the trend reverses direction.  We tested the significance of the trend
by fitting the data with five models: a constant phase model, a linear model, a bilinear
model, a bilinear model with the slope of the second line segment
fixed to 0, and a bilinear model with both the slope and intercept of
the second line segment fixed to 0.  We obtain a BIC of 185, 168, 136, 145,
and 186 respectively.  Thus, the reversal at $3400 K$, despite being based on
only three data points, is significant from a purely statistical
perspective.  The fit results for the first line segment are $b_1 =
47.7 \degree \pm 4.7 \degree$, $m_1 = -0.014 \pm 0.002$ deg/K; for the crossover
point, $T_c=3410 \pm 110$; and for the second line segment, $m_2 =
0.039_{-0.011}^{+0.017}$ deg/K.

Despite the statistical significance, the rise is still based
only three planets, and may not be real.  The test assumes
Gaussian errors, while the actual errors are in reality both
asymetric and non-Gaussian.  Even more importantly, \spitzer light
curves are notorious for having bizarre and unexplained instrumental systematics
which can affect the fitted parameters in ways that are subtle and
difficult to diagnose.  In
this paper alone, we have seen this for HD 149026b in the 3.6 \um
band, and WASP-12b for two different observations in the 3.6 \um
band.  Other examples of problematic behavior include \cite{stevenson_2017},
where two separate visits in the same band resulted in very different nightside
fluxes.  A sprinkling of unmodelled systematics, plus a smattering of
bad luck, could be sufficient to destroy the final rise in
temperature.

On the other hand, there are physical reasons to be less skeptical.
First, although the reversal at $3400 K$ has never been predicted or
previously noted, the initial drop is unsurprising--phase offsets are expected to decrease with increasing
temperature because the radiative timescale drops steeply with
temperature \citep{perez_becker_showman,komacek_showman}.  Second, it is also clear that some kind of break must occur at $3400 K$--if
the downward trend continued, the phase offset would become westward
at higher temperatures, which is physically implausible.

The only significant outlier in this trend is the 3.6 \um observation for HD 149026b, which we discuss in \S\ref{subsec:hd149026b_ch1}.  Additionally, as discussed in \S\ref{subsec:comparison}, \cite{wasp12b_phase_curve} presented two contradictory sets of results for WASP-12b's phase curve,
based on two distinct analysis methods.  Had we used the other version, the 4.5 \um
phase curve offset would have been nearly identical, but the 3.6 \um phase curve
offset would be at a physically implausible 53$\degr$--another clear
outlier.  The fact that both potential outliers in this plot are based on problematic data sets gives confidence to the reality of the trend.  The trend is even more striking when one considers that it is between
two relatively reliably measured quantities.  The irradiation temperature is
dependent only on the stellar effective temperature and
$a/R_*$, both of which are easily measured.  The phase offset is harder
to measure (e.g., \S\ref{subsec:comparison}), but unlike albedos and efficiencies, it is a purely
empirical quantity.

To understand this trend, we took previously published SPARC GCM
simulations \citep{kataria_2016} and plotted phase offset against irradiation
temperature in Figure \ref{fig:model_T_vs_offset}.  The sample of
planets in these GCMs is different from those presented in the
observations.  All models are at solar metallicity and have no TiO, in
order to ensure a more uniform comparison.  As expected, the predicted
phase offsets from these models decrease with increasing irradiation
temperature until approximately 3000 K.  The offsets decrease at a
rate of $-0.017$ deg/K.  However, instead of rising
towards the highest temperatures they instead plateau around a minimum
phase offset of approximately 10$\degr$.  It is also worth noting that
in these models the 3.6 \um bandpass has a larger phase shift (by an
average of 4.1$\degr$) for all
but the coolest planet, indicating that this wavelength probes deeper
into the atmosphere.  In our observations, the 4.5 \um phase curves
have larger phase shifts (and thus deeper photospheres) than the
3.6 \um phase curves for every planet except the coldest, the average
difference being 6.2 degrees.  The
universality of this trend among our relatively diverse sample of hot
Jupiters is suggestive, and should help guide future modeling
efforts in this area.

There are a combination of factors that may or may not explain the
discrepancy between observations and GCMs. For example, the
addition of high altitude clouds to the GCMs could help by
decreasing the dayside photospheric pressure, which would
systematically reduce the size of the predicted phase offsets and
provide a better match to the observational data.  Such clouds appear
to provide a good match to the optical phase curve offsets measured
for the hot Jupiters located in the Kepler field
\citep{parmentier_2016, demory_2013,shporer_2015,angerhausen_2014}
and have also been postulated to explain other infrared phase-curve
observations \citep[e.g.][]{kataria_2015,stevenson_2017}.  However, \cite{roman_2017} complicate this explanation by showing that clouds do not always lead to lower phase offsets.  They find that in the case of Kepler-7b, inhomogenous clouds distributed along the western terminator result in a higher phase offset, while global clouds result in a marginally lower offset, compared to the clear atmosphere case.  A
super-solar metallicity atmosphere could also provide a similar
effect; enhanced metallicity results in enhanced opacities, such that
the photosphere is higher in the atmosphere where the day-night contrast is larger and the phase offset is smaller (e.g., \citealt{kataria_2015}). At high temperatures, the presence of dayside
temperature inversions produced by gas-phase TiO/VO might also affect the predicted phase offsets, as this will change the opacity of the atmosphere and hence what altitudes are probed (e.g., \citealt{sparc_model}). 

MHD effects such as Lorentz drag and Ohmic dissipation are also likely
to be important (e.g.,
\citealt{perna_2010,batygin_2010,menou_2012,rauscher_2012,ginzburg_2016}).
Because
Lorentz drag and Ohmic dissipation are two facets of the same underlying processes, we can turn to the literature on Ohmic dissipation and radius inflation in hot Jupiters to determine the regime in which these effects become important.  For hot Jupiters with an appreciable magnetic field, the effect of this magnetic field on the atmospheric circulation will depend on the ionization fraction of the planet's upper atmosphere.  Previous models have concluded that alkali metals such as Na and K will provide the dominant source of ions in these atmospheres (e.g., \citealt{batygin_2010}).  As the ionization fraction increases the strength of the Lorentz drag and also the amount of radius inflation due to Ohmic dissipation increase as well, resulting in a peak radius inflation at equilibrium temperatures of around 1500 K.  At higher temperatures, the atmospheric circulation is effectively suppressed by the magnetic drag and less energy is deposited in the planet's interior.  \cite{thorngren_2017} find compelling evidence for a peak in the radius inflation of hot Jupiters around 1500 K and a decline thereafter, in good agreement with these models.  With this picture in mind, the addition of Lorentz drag around 1500 K is likely the explanation for why the observed phase offsets at higher temperatures decrease to values consistent with zero, while the GCMs predict a minimum phase offset around 10$\degr$.  However, this simple picture appears to conflict with the observed increase in phase offset for the most highly irradiated planets, as the amount of Lorentz drag should remain constant over this highly irradiated regime.  Although there has been some recent work on atmospheric circulation in the MHD-dominated regime (e.g., \citealt{rogers_2017}), it is not yet clear whether more careful modeling can reproduce the observed trend in phase offsets at these temperatures.    

As a last point, we consider possible explanations for the relative offsets observed between the two \emph{Spitzer} bands.  As noted earlier, the measured 4.5 \um offset is consistent larger than the measured 3.6 \um offset for every planet except for HD 189733b, but the opposite is consistently true in the model predictions for these planets.  Atmospheric chemistry would seem to be the obvious explanation:  \methane is a major absorber within the 3.6 \um bandpass, while CO is a strong absorber within the 4.5 \um bandpass, so the relative abundances of these two molecules could easily shift the relative photospheric pressures in these two bands. Increasing the amount of \methane via vertical mixing or other disequilibrium chemistry processes would increase the opacity and decrease the photospheric pressure in the 3.6 \um band, resulting in a larger day-night contrast and smaller phase offset. Similarly, decreasing the amount of CO in the atmosphere would decrease the opacity in the 4.5 \um band, shifting the photosphere to higher pressures with a smaller day-night temperature contrast and a larger phase offset.  Still, an enhanced methane abundance would require a drastic departure
from equilibrium chemistry, as CO is expected to be the major
carbon-bearing molecule at temperatures relevant to hot Jupiters.
\cite{review}, for example, show that the delineation between CO and
\methane dominance at 1500 K is at 10 bars, rising to 100 bars at 2000
K. For hot planets the abundance of \methane at photospheric pressures
(approx. 100 mbar) should be tiny. Even for HD 189733b, the coldest planet in
Figure 14, the \methane abundance is nearly 3 orders of magnitude below the
CO abundance for any reasonable photospheric pressure, as shown in
Figure 3 of \cite{review}.

\section{Conclusions}
\label{sec:conclusions}
In this paper we present new phase curve observations for WASP-33b and HD 149026b at 3.6 \um and 4.5 \um.  Our measured parameters are in good agreement with previously published transit and secondary eclipse observations of these two planets, and we use our new phase curve observations to investigate the atmospheric circulation patterns of these two planets.  We use a simple toy model to estimate the brightness temperatures, albedo, and recirculation efficiency of both planets under the assumption that they emit as blackbodies and find that WASP-33b appears generally similar to other hot Jupiters despite its unusually high irradiation level.  On the other hand, HD 149026b has a typical recirculation
efficiency but an albedo of 0.6--the highest ever measured.  This
albedo strongly suggests the presence of clouds, which could easily be
confirmed with \emph{HST} transmission spectroscopy.  Intriguingly, we
find strong evidence for a correlation between the masses of planets
with published thermal phase curves and their inferred albedos; this
may be indicative of the role that surface gravity plays in the
settling of cloud particles.

We also compared our measured phase curves for these two planets to predictions from GCMs.  For HD 149026b,
we considered models with $1\times$ and $30\times$ solar metallicity, both of which provided an unusually poor match to the observed phase curve shapes.  Based on this planet's high inferred albedo and enhanced bulk metallicity, it seems likely that even higher metallicity GCMs incorporating clouds could provide a better match to these data. For WASP-33b, we considered models with and without TiO; although there were still some discrepancies, these models were overall a much better fit than in the case of HD 149026b.  We note that MHD effects likely dominate the atmospheric circulation for highly irradiated planets like WASP-33b, and present an obvious avenue for future investigations.

Lastly, we placed these two planets in context by comparing
their observed phase offsets in each band to those of other planets.
We find a strong correlation between measured phase
offset and irradiation temperature, where the observed
offset decreases with increasing irradiation temperature to a minimum
around 3400 K, and then rises again for the most highly irradiated
planets.  Although this decreasing trend with increasing irradiation
is predicted by GCM simulations of these planets, in practice the size of
the observed phase offsets for the coolest planets appear to be lower
than predicted by the GCMs.  We propose that this can be explained by
the presence of high altitude cloud layers in these atmospheres, which
decrease the photospheric pressure probed in these two bands.  At
higher temperatures, we find that the observed phase offsets decrease
to zero for irradiation temperatures near 3400 K, while the GCMs
predict a minimum phase offset of 10$\degr$ for planets in this
temperature regime.  We propose that this discrepancy can be resolved
by the inclusion of MHD effects such as Lorentz drag, which would
serve to further reduce the speed of atmospheric winds and decrease
the size of the observed phase offset.  We note that the trend of
increasing phase offset with increasing temperature for the most
highly irradiated planets is not well-matched by this simple picture,
but perhaps could be explained with more sophisticated circulation
models incorporating the full range of MHD effects.  Finally, we
propose that the relative phase offsets at 3.6 and 4.5 \um, which are
consistently the opposite of those predicted in the GCMs, might
be explained by a change in the assumed atmospheric compositions
and/or chemistries of these planets.  

\section{Acknowledgments}
This work is based on observations made with the \emph{Spitzer Space Telescope}, which is operated by the Jet Propulsion Laboratory, California Institute of Technology under a contract with NASA. Support for this work was provided by NASA through an award issued by JPL/Caltech. H.A.K. acknowledges support from the Sloan Foundation.

\bibliographystyle{apj} \bibliography{ms}

\end{document}